\documentclass[aps,pra,reprint,superscriptaddress,floatfix]{revtex4-1}
\usepackage{braket}
\usepackage{amsmath}
\usepackage{graphicx}
\usepackage{placeins}
\usepackage{xcolor}
\usepackage{stackengine}
\graphicspath{{./figures/}}
\usepackage[bookmarks=false]{hyperref}
\hypersetup{
    pdftitle={Self-Organization of Atoms Coupled to a Chiral 1D Reservoir},
    pdfauthor={Zachary Eldredge},
    colorlinks=true,            
    pdfstartview=Fit,           
    pdfpagemode=UseOutlines,    
    pdfpagelayout=TwoPageRight
}

\newcommand{\gamtot}{\Gamma_\mathrm{tot}}
\newcommand{\gamoned}{\Gamma_\mathrm{1D}}

\newcommand{\see}[1]{\sigma_{ee}^{#1}}
\newcommand{\sgg}[1]{\sigma_{gg}^{#1}}
\newcommand{\seg}[1]{\sigma_{eg}^{#1}}
\newcommand{\sge}[1]{\sigma_{ge}^{#1}}
\newcommand{\dd}[1]{\mathrm{d} #1}
\newcommand{\abs}[1]{\ensuremath{\left| #1 \right|}}
\newcommand{\half}{\frac{1}{2}}
\newcommand{\pdpd}[2]{\frac{\partial #1}{\partial #2}}
\newcommand{\expect}[1]{\langle #1 \rangle}
\newcommand{\figref}[1]{Fig.~\ref{#1}}

\newcommand{\nnb}{\nonumber \\}

\begin{document}
\title{Self-organization of atoms coupled to a chiral reservoir}
\date{\today}
\author{Zachary Eldredge}
\affiliation{Joint Quantum Institute, NIST/University of Maryland, College Park, Maryland 20742, USA}
\affiliation{Joint Center for Quantum Information and Computer Science, NIST/University of Maryland, College Park, Maryland 20742, USA}
\author{Pablo Solano}
\affiliation{Joint Quantum Institute, NIST/University of Maryland, College Park, Maryland 20742, USA}
\author{Darrick Chang}
\affiliation{ICFO - Institut de Ciencies Fotoniques, The Barcelona Institute of Science and Technology, 08860 Castelldefels (Barcelona), Spain}
\author{Alexey V. Gorshkov}
\affiliation{Joint Quantum Institute, NIST/University of Maryland, College Park, Maryland 20742, USA}
\affiliation{Joint Center for Quantum Information and Computer Science, NIST/University of Maryland, College Park, Maryland 20742, USA}

\begin{abstract}
Tightly confined modes of light, as in optical nanofibers or photonic crystal waveguides, can lead to large optical coupling in atomic systems, which mediates long-range interactions between atoms. These one-dimensional systems can naturally possess couplings that are asymmetric between modes propagating in different directions. Strong long-range interaction among atoms via these modes can drive them to a self-organized periodic distribution. In this paper, we examine the self-organizing behavior of atoms in one dimension coupled to a chiral reservoir. We determine the solution to the equations of motion in different parameter regimes, relative to both the detuning of the pump laser that initializes the atomic dipole-dipole interactions and the degree of reservoir chirality. In addition, we calculate possible experimental signatures such as reflectivity from self-organized atoms and motional sidebands.
\end{abstract}

\maketitle

\section{Introduction}
\label{sec:introduction}
Optical systems such as tapered optical nanofibers and photonic crystal waveguides have attracted great interest due to their ability to support tightly confined modes of light \cite{Brambilla2010,Nayak2013,Ravets2013,Morrissey2013}. 
The small mode volume in such setups results in a large coupling between the guided mode and atoms near the structure, modifying the optical properties and responses of nearby atoms \cite{Vetsch2010,Meng2015,Reitz2013,Goban2015,Rephaeli2011,Arcari2014}. 
Atoms can be held in the evanescent field along a nanofiber by use of a two-color dipole trap, which provides confinement in azimuthal and radial directions. By introducing a standing wave, the trap can be made to constrain the atoms along the axial direction \cite{Goban2012,Lacroute2012,Lee2015,Schneeweiss2014,LeKien2013,Vetsch2012}. In photonic-crystal waveguides, a trap scheme making use of Casimir-Polder forces could accomplish a similar goal \cite{Hung2013,Gullans2012}. If there is no axial trapping, the atoms are free to move along the fiber or waveguide. The atoms can then be driven either through the guided mode or by side illumination. If dissipation is introduced, then self-organized solutions to the equations of motion with stable spatial configurations of the atoms exist \cite{Chang2013}. Similar phenomena have also been theoretically explored \cite{Domokos2002} and observed \cite{Black2003} for cold atoms in a cavity.
This organization arises due to the forces associated with emission and re-absorption of photons into the guided modes, which can be described by an effective dipole-dipole interaction \cite{Gonzalez-Tudela2015,Goban2014,Yu2014,Sague2007,Shahmoon2014a,Shahmoon2014b}. Furthermore, the effectively infinite-range character of photon-mediated exchange in 1D and strong coupling to the guided modes causes these inter-atomic forces to become quite prominent \cite{Shahmoon2014}.  The final configuration of the atoms in position and momentum space will depend on optical properties of the driving field.

The optical interface in nanophotonic systems need not be perfectly bidirectional \cite{Mitsch2014a,Scheel2015a,LeKien2014,Sollner2015}. This is due to the fact that a tightly confined mode has longitudinal polarization \cite{Junge2013}. An atom may preferentially interact with the left- or right-propagating mode, which could provide valuable tools for quantum optics \cite{Metelmann2015a}.
However, self-organized configurations may be modified or nonexistent in cases where the interaction has this chiral nature. Some study for small numbers of atoms has been done in a scattering-matrix formulation in Ref.~\cite{Holzmann2014}, but chiral systems were not the main focus of that work. 
Self-organized configurations have several potential technological applications, including optical processing, nanophotonic interfaces, and quantum information storage \cite{Gouraud2015,Sayrin2015a,Muschik2011,Mahmoodian2016}. In addition, they may serve as simulators for various quantum mechanical many-body models \cite{Douglas2015}. In addition, the strong optical nonlinearity due to the atoms coupled to the waveguide can lead to photon-photon interactions \cite{Douglas2015a,Firstenberg2013,Shahmoon2011,Shahmoon2016}.

In this paper, we present analytic and numerical results showing that chiral systems will still produce self-organized solutions. We will show how these solutions differ from the symmetric case when the pump field is both near and far from resonance and present proposals for experimental detection of self-organization behavior in chiral systems that include an analysis of optical response to probe lasers as well as an examination of the structure of motional sidebands in chiral systems.

The remainder of this paper is organized as follows. In Sec.~\ref{sec:setup}, we introduce the system under investigation. We then, in Sec.~\ref{sec:purelychiral}, examine cases where the interaction is entirely unidirectional both near and far from resonance before examining (Section \ref{sec:arbitrarychiralities}) the behavior when the system is biased toward one direction but still allows atoms to couple to both left- and right-propagating modes with different strengths. Next we discuss in Sec.~\ref{sec:experimentalconsiderations} means of producing couplings with varying chirality and identifying experimental signatures of differently self-organized configurations. Finally, Section \ref{sec:conclusion} summarizes our results, and the Appendix contains mathematical details omitted from the main text.
\section{Setup}
\label{sec:setup}
\subsection{Physical System and Equations of Motion}
\begin{figure}[tb]
	\includegraphics[width=8.6cm]{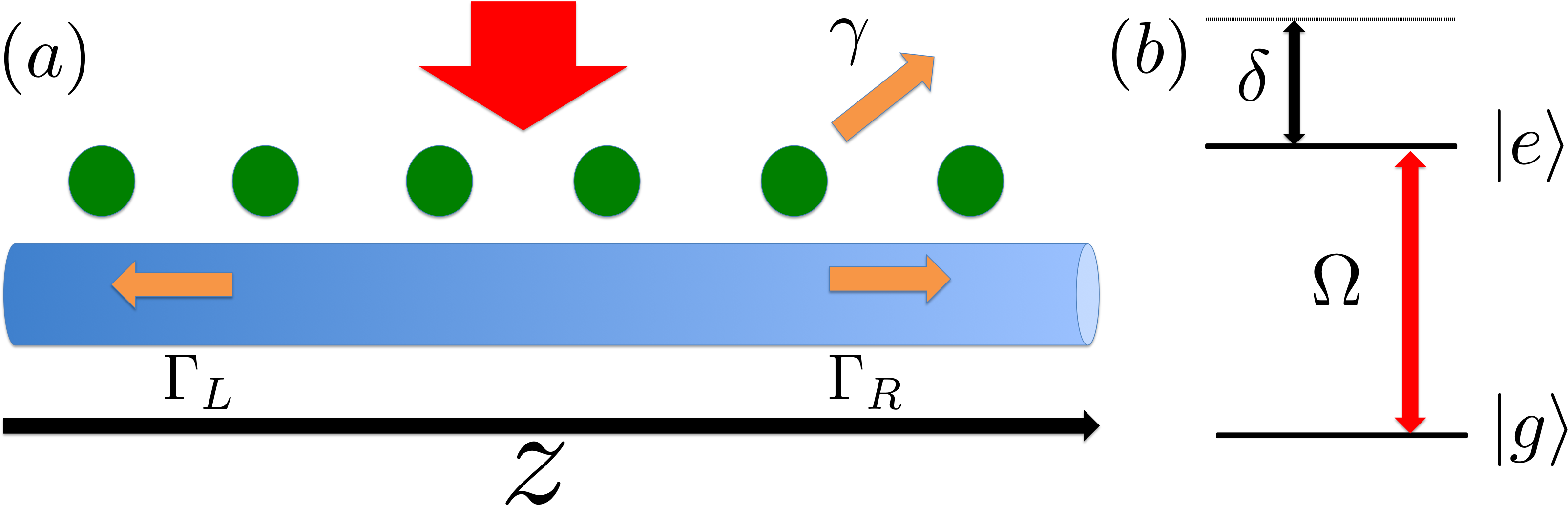}
	\caption{ (a) Atoms are free to move along the axis $z$ and are illuminated from the side by a pump field of real Rabi frequency $\Omega$ and detuning $\delta$. They decay with different rates into left- and right-propagating guided modes ($\Gamma_L$ and $\Gamma_R$) and into free space ($\gamma$). (b) The level diagram of the atom in question and the applied field. Although drawn here as positive (blue-detuned), $\delta$ may be of either sign.}
	\label{fig:setup}
\end{figure}

Suppose that $N$ atoms are free to move axially along the waveguide, and that the confinement in the radial and azimuthal directions is strong enough to neglect these degrees of freedom, making the system effectively one-dimensional. 
The atoms have a single transition with angular frequency $\omega$ and associated wavevector in the fiber $k$. The ground and excited state of the atoms are denoted $\ket{g}$ and $\ket{e}$, respectively. The atoms are illuminated from the side by a pump laser with frequency $\omega_L$, corresponding detuning $\delta = \omega_L - \omega$, and real Rabi frequency $\Omega$. It should be noted that it is not necessary to illuminate from the side, as atoms can also be illuminated through the guided mode. These dynamics would be different as each atom would be driven at a different phase. While this may form an interesting basis for future work, here we restrict ourselves to side illumination. However, the numerical methods of this work would also be applicable to these systems if the appropriate phase was added to the driving Rabi frequencies. This setup is illustrated in \figref{fig:setup}.
The total decay rate for an atom into all modes, both guided and free space, is denoted $\gamtot$. Some of this decay goes into free space and is lost to the system, which occurs at rate $\gamma$. The decay rate specifically into the guided modes is $\gamoned$, a sum of $\Gamma_L$ and $\Gamma_R$, the decay into left and right directions, respectively:
\begin{equation}
	\gamtot = \gamoned + \gamma = \Gamma_L + \Gamma_R + \gamma 
\end{equation}
In numerical results of this paper, $\gamoned = 0.25 \gamtot$, which is a value easily accessible in current photonic crystal waveguides \cite{Yu2014}. This was the value used in previous work on the symmetric case, so that all results will match those of Ref.~\cite{Chang2013} in the limit $\Gamma_L = \Gamma_R$. While $\gamoned = 0.25 \gamtot$ is higher than accessible with current optical nanofiber technology \cite{Corzo2016}, reducing $\gamoned$ does not qualitatively change the self-organization behavior, as discussed in Sec.\ \ref{sec:experimentalconsiderations}.  In particular, self-organized stable configurations  still exist if $\gamoned$ takes a smaller or larger value. Higher values of $\gamoned/\gamtot$ have the advantage of stronger self-organized potentials, and therefore require less cooling, also discussed in Sec.~\ref{sec:experimentalconsiderations}. 

We define a ``chiral decay rate''
\begin{equation}
	\chi \equiv \Gamma_R - \Gamma_L.
	\label{eq:chidef}
\end{equation}
The advantage of separating off the chiral decay rate is that the equations of motion will now be similar to previous work \cite{Chang2013} with an additional term that represents the excess coupling in one direction, as will be shown in Eqs.~\eqref{eqn:eomz} - \eqref{eqn:eomp}. The chirality of the system can be characterized by the fraction of the decay into guided modes that is preferentially directed in one direction. We define $\chi_r \in \left[-1 , 1 \right]$,
\begin{equation}
	\chi_r \equiv \frac{\chi}{\gamoned} = \frac{\Gamma_R - \Gamma_L}{\gamoned}.
\end{equation}
If $\chi_r = 0$, then $\Gamma_R = \Gamma_L$, whereas if $\chi_r = 1$, then $\Gamma_L = 0$ and all spontaneous emission into the fiber is into rightward-propagating modes. In this fully-chiral case, the interaction is one-way, and atoms only affect atoms to their right. In this work, we without loss of generality take $\chi_r \geq 0$. 
In all cases, numerical parameters are those of the cesium $\mathrm{D}_2$ line \cite{Steck2010b}: the transition has wavenumber in vacuum $\hbar k  / (2 \pi) = 11727 \ \mathrm{cm}^{-1} $ and decay rate in free space $\gamma = 5.2 \ \mathrm{MHz}$. Throughout this paper we present results as dimensionless ratios wherever possible, but atomic parameters are necessary in our simulations to specify the ratio between the recoil frequency $\omega_r = \hbar k^2 / 2 m$ (where $m$ is atomic mass) and total emission rate $\gamma$.

The full atom-field Hamiltonian can be reduced to a dissipative spin model by integrating out photonic degrees of freedom, as in Ref.~\cite{Chang2012}. This procedure is carried out explicitly in the Appendix\ref{app:reduce}. The evolution of the system density matrix $\rho$ is then given by the master equation
\begin{equation} \label{eqn:master}
\dot{\rho} = -\frac{i}{\hbar} \left[H, \rho \right] - \half \sum_m \left(L_m^\dag L_m \rho + \rho L_m^\dag L_m - 2 L_m \rho L_m^\dag \right).
\end{equation}
This can be compared to previous work on symmetry-broken master equations, Refs.~\cite{Carmichael1993,Gardiner1993}. The coherent evolution is described by the following Hamiltonian,
\begin{align}
	\label{eqn:hamiltonian}
	H &= \sum_j \left[ \frac{p_j^2}{2m} -\hbar \delta \see{j} - \hbar \left( \Omega \seg{j} + h.c. \right) \right] \nonumber \\ &+ \sum_j \sum_{i \neq j} \hbar \Gamma_L \sin \left( k \left| z_i - z_j \right| \right) \seg{i} \sge{j} \nonumber \\&+ i \hbar \frac{\chi}{2} \sum_j \sum_{i<j} \left[ \seg{j} \sge{i} e^{-i k \abs{z_j - z_i}} - \seg{i}\sge{j} e^{i k\abs{z_j - z_i}} \right],
\end{align}
where $\sigma_{\mu \nu}^j$ is the atomic state operator $\ket{\mu} \bra{\nu}$ for the atom $j$ with $\mu, \nu\in \left\{ e, g \right\}$ corresponding to excited and ground states. The position of the $j$th atom is given by $z_j$. 

The first term of Eq.~\eqref{eqn:hamiltonian} is a kinetic energy for momentum $p_j$ and mass $m$. The second term, proportional to $\delta$, is the energy of the excited state in the rotating frame.
 The third term, proportional to $\Omega$, represents the interaction between the pump laser and the atoms. The fourth term, proportional to $\Gamma_L$, sums over all pairs and describes the process of exchange of atomic excitations, mediated by a guided photon in the case of a symmetric coupling.  The sinusoidal modulation reflects the propagation phase of the photon mediating the interaction; the interaction is infinite-ranged.

 The excess in the coupling to the other mode has been moved into the fifth term, proportional to $\chi$. We have chosen to work with $\Gamma_L$ and $\chi$ (rather than $\Gamma_\mathrm{L,R}$) in Eq.~\eqref{eqn:hamiltonian} to facilitate comparison to the symmetric case, simply by setting $\chi = 0$. 

The incoherent evolution in the master equation [Eq.\ (\ref{eqn:master})] is described by the jump operators $L_m$, where $m$ runs over the $N$ independent atomic emissions into free space with rate $\gamma$, the symmetric decays in both directions of the guided mode with rate $\Gamma_L$, and the asymmetric decay with rate $\chi$:
\begin{equation}
	\label{eqn:3Ddecay}
	L_j = \sqrt{\gamma} \sge{j},
\end{equation}
\begin{equation}
	\label{eqn:LRdecay}
	L_{L,R} = \sqrt{\Gamma_{L}} \sum_j e^{\mp i k z_j} \sge{j},
\end{equation}
\begin{equation}
	\label{eqn:chidecay}
L_{\chi} = \sqrt{\chi} \sum_j e^{-i k z_j} \sge{j}.
\end{equation}
The exponential in Eq.~\eqref{eqn:LRdecay} has a minus sign for the decays into rightward-propagating modes. This formulation is equivalent to previous work in chiral dissipative systems \cite{Ramos2014}, where $\Gamma_L$ and $\Gamma_R$ are kept separated. In the $\chi_r = 1$ case, this is a cascaded system, as described by Refs.~\cite{Carmichael1993,Gardiner1993,Stannigel2012}. The jump operators $L_{i,L,R}$ are identical to those used to describe atoms couples symmetrically to a waveguide, as in Ref.~\cite{Chang2013}.

We work in a classical limit for the motion of the atoms, obtaining evolution equations for the expectation values of momentum and position. This approximation is justified if the atoms are not cooled to near their motional ground state.
As we will be dealing exclusively with expectation values from now on, we will omit angle brackets for notational convenience--$z$, $\sigma_{ge}$, and $p$ should be read as expectation values. In addition, we will make the approximation that atomic excitation is low enough that saturation can be neglected (we treat a factor $\sigma_{gg} - \sigma_{ee} \approx 1$), which also implies that the spins are not correlated, i.e., $\expect{ \seg{j} \sge{i}} \approx \expect{\seg{j}} \expect{\sge{i}}$. The de-correlation follows by writing a two-atom wavefunction and evaluating the two expectation values. Since $\sigma^{j}_{ee}$ and $\sigma_{ee}^{i}$ are both small, the $\ket{ee}$ component of the wavefunction can be neglected, and under this condition the two expectation values are equal. The equations of motion are:
\begin{align}
	\label{eqn:eomz}
	\dot{z}_j &= \frac{p_j}{m}, \\
	\label{eqn:eoms}
	\dot{\sigma}^j_{ge} &= \left( i \delta - \frac{\gamtot}{2} \right) \sge{j} + i \Omega - \Gamma_L \sum_{i \neq j} \sge{i} e^{i k \abs{z_j - z_i}} \nonumber \\&- \chi \sum_{i < j} \sge{i} e^{i k \abs{z_j - z_i}}, \\
	\label{eqn:eomp}
	\dot{p}_j &= - \hbar k \chi \abs{\seg{j}}^2 -2 \hbar k \chi  \Re \sum_{i<j} \seg{j} \sge{i} e^{i k \left(z_j - z_i \right)}\nonumber \\& - 2 \hbar k \Gamma_L \Re \sum_i \left[ \sge{i} \seg{j} e^{i k \abs{z_j - z_i} }  \mathrm{sgn} \left(z_j - z_i \right) \right] -\gamma_p p_i .
\end{align}

The term $\gamma_p$ does not arise from Eq.~\eqref{eqn:master}; it is an external damping which removes energy from the system so that the atoms can come to equilibrium. This is required to achieve numerical convergence; without this term the atoms do not achieved a self-organized equilibrium. The external damping in these equations could be provided by the same cooling systems that will be necessary to trap the atoms initially. A standard optical molasses setup would destroy the coherences that are necessary to produce the dipole-dipole interaction. However, other cooling schemes may be able to provide the needed damping force. One possibility could be a form of stroboscopic cooling, in which the system is alternated between periods of cooling and periods of interaction. The resulting dynamics would then resemble those analyzed here. A further discussion of cooling and the damping coefficient required to stabilize the system can be found in Sec.~\ref{ssec:other}. 

The momentum equation, Eq.~\eqref{eqn:eomp}, contains a term $-\hbar k \chi \abs{\sge{i}}^2$. This is not present in the symmetric case (where $\chi = 0$). It arises because spontaneous emission is no longer symmetric. The symmetric case lacks this term and converges to states with $p_j = 0$ for all $j$, as the $\gamma_p$ damping term removes kinetic energy from the system, but this is not expected for chiral couplings. Instead, self-organized solutions in the chiral case will be those for which $p_i = p_j$ for all pairs of atoms $i, j$. There will be a center-of-mass motion but the atoms will remain fixed in relative spacing. The final momentum will be inversely proportional to $\gamma_p$, and without damping the system will continue accelerating.

\subsection{Numerical Strategy For Finding Steady-State Behavior}
\label{sec:numerics}
We begin with atoms in random positions and with no initial momentum or coherences. At time $t = 0$, a pump field of real Rabi frequency $\Omega$ and detuning $\delta$ switches on, and atoms begin interacting. The system is evolved in time according to Eqs.~\eqref{eqn:eomz}-\eqref{eqn:eoms} until all the derivatives in the equations of motion are zero, except for position (due to the expected center-of-mass motion). 

In general, there may not be a unique steady state that arises from this procedure. First, since the atomic positions only enter into the equations of motion through phase factors $e^{ikz}$, there is a multiplicity of solutions arising from invariance under shifts of the atomic position by integer wavelengths. In addition, different initial conditions in the random positions of the atoms could in principle lead to different steady-state solutions, which is particularly true when the atoms are driven close to resonance. In order to find a replicable and reliable steady state, we follow a particular route in parameter space. 
As discussed in Sec.~\ref{sec:ws}, for infinite detuning the atomic coherences de-couple from the positions, which leads to a unique steady-state solution. In practice, the procedure starts with a detuning $\delta = \pm 20 \gamtot$, large enough that a deterministic configuration emerges close to the $\delta = \pm \infty$ solution.
After the initial run, the $\delta = \pm 20 \gamtot$ equilibrium results for $p$, $\sigma$, and $z$ are used as initial conditions for a new simulation with $\delta$ slightly closer to resonance. This scheme can find steady states under a near-resonant pumping that can be replicated from one run of the program to the next. In addition, such a preparation procedure could perhaps mirror an experimental scheme to produce these arrangements. 

This procedure illustrates the necessity of the damping rate $\gamma_p$. If the initial condition is far away from the steady-state, then there will be potential energy stored in the initial positions of the atoms. Since atomic interactions mediated by photon exchange conserve momentum, the self-organizing forces are not capable of bringing the configuration to a low-energy equilibrium; dissipation is required to achieve self-organization. However, the final configuration is determined by interatomic distances, and these are not affected by the damping rate as can be seen by setting $\dot{p}_j = 0$ in Eq.~\eqref{eqn:eomp}. Therefore, the damping rate does not affect what steady-state configuration is found. If $\gamma_p$ is set to 0 after a steady state has been achieved, the self-organized configuration will remain intact, but it will now accelerate due to asymmetric spontaneous emission. The self-organization survives in an accelerating frame. 

\section{Chiral Self-Organization}
\label{sec:purelychiral}
In this section we explore the types of self-organized solutions that arise when $\chi_r = 1$, that is, in the case of a one-way (cascaded) interaction. In this regime, the asymmetric decay rate $\chi$ is equal to the total decay into guided modes $\gamoned$. However, to make clear which terms we are including from the general equations, we still write $\chi$. First, we discuss the simpler case of the ``weak-scattering'' limit (large detuning) before deriving a general solution which allows for any detuning.
\subsection{Weak-Scattering}
\label{sec:ws}
In Eq.~\eqref{eqn:eoms}, one sees that the atomic coherence of each atom $j$ is driven by a combination of the external field with Rabi frequency $\Omega$ and the fields rescattered into the waveguide by other atoms $i \neq j$. The weak-scattering limit is defined as the regime in which the re-scattered fields are negligible. Quantitatively, a sufficient condition is $N \gamoned \ll \sqrt{\delta^2 + \gamtot^2/4}$ and $\Omega \ll \gamtot , \delta$. This allows us to say that $\sge{j} =  s_0$, for all $j$ which can be determined by solving the $\sge{j}$ equation \eqref{eqn:eoms} if the terms involving other atoms are discarded:
\begin{equation}
	\abs{s_0}^2 = \frac{\abs{\Omega}^2}{\delta^2 + \gamtot^2/4}.
\end{equation}
The value of $s_0$ actually has no effect on the steady-state positions of the atoms in the weak scattering case. Throughout this paper in the numerics we set $\Omega = .1 \sqrt{\delta^2 + \gamtot^2/4}$ so as to enforce constant saturation of the atoms even as $\delta$ changes. Note that a different prefactor value for $\Omega$ rescales the $\sigma$ expectation values, and choosing different values of $\Omega$  only affects the transient behavior but not the actual steady-state solutions. 

The dynamics of a chiral system are not Hamiltonian. This is because every atom has a damping term included which does not conserve energy. This differs from the symmetric case, where the damping can be omitted without leading to runaway momentum and the problem can be solved by energy minimization of the many-body potential in the weak-scattering limit. We use the chiral part of the equation of motion for momentum, Eq.~\eqref{eqn:eomp}, but we set all $\sge{j} = s_0$ due to the weak-scattering condition:
\begin{equation}
	\dot{p}_j = - \hbar k  \chi s_0^2 -2 \hbar k \chi s_0^2 \Re \sum_{i<j} e^{i k \left(z_j - z_i \right)} - \gamma_p p_j.
\end{equation}
For an equilibrium state, the momentum will be constant, so
\begin{equation}
	p_j = - \frac{\hbar k \chi s_0^2}{\gamma_p} \left[1  -2 \Re \sum_{i<j} e^{i k \left(z_j - z_i \right)} \right].
	\label{eqn:constpj}
\end{equation}
For the leftmost atom, the second term disappears as an empty sum. The right-hand side of Eq.~\eqref{eqn:constpj} must be the same for every atom to yield a self-organized configuration, $p_i = p_j$ for all $i, j$. Therefore the sum must vanish for all atoms. The condition is
\begin{align}
\label{eqn:energycondition}
	\Re \sum_{i<j} e^{i k \left(z_j - z_i \right)} \propto \pdpd{ }{z_j} \sum_{i<j} \sin k \left( z_j - z_i \right) = 0.
\end{align}
The condition of stationary momentum is identical to a condition in which a potential acting on a single particle is minimized,
\begin{equation}
	\label{eqn:singleparticlepotential}
	V_j = 2 \hbar \chi s_0^2 \sum_{i < j} \sin k \left(z_j - z_i \right).
\end{equation}	
This resembles a dipole potential acting on the $j$th atom which considers only the influence of atoms to its left. Minimizing this ensures that the sum in Eq.~\eqref{eqn:energycondition} vanishes. Therefore, equilibrium for each particle is achieved by minimizing this potential, even though the system as a whole does not minimize any general many-body potential. 

As previously noted, there is a multitude of solutions associated with the displacement of atoms by an integer number of wavelengths. Therefore, instead of the actual position $z_j$, we will consider the fractional distance $f_j$ satisfying $ 0< f_j \leq 1$, defined such that $z_j = \lambda( n_j + f_j)$ where $n_j$ is an integer \cite{Chang2013}. All arguments $k z_j$ to periodic functions are replaced with $2 \pi f_j$.  In addition, the position of the leftmost atom is defined to be $f_0 = 1$ without loss of generality. The position of the next atom is found by minimizing the next potential, which is
\begin{equation}
V_1 = 2 \hbar \chi s_0^2 \sin 2 \pi \left(f_1 - 1 \right) = 2 \hbar \chi s_0^2 \sin 2 \pi f_1.
\end{equation}
This is minimized by setting $f_1 = 3/4$. The potential experienced by the next atom follows from these positions, and so on. To find $f_j$, we rewrite Eq.~\eqref{eqn:singleparticlepotential} as
\begin{equation}
\label{eqn:sumofsines}
V_j = E_j \cos \left(2 \pi f_j - \delta_j\right),
\end{equation}
where 
\begin{align}
	\delta_j &= \arctan  \left( \frac{ \sum_{i < j} \sin 2 \pi f_i}{\sum_{i<j} \cos 2 \pi f_i} \right) - \frac{\pi}{2},  \\
\label{eqn:recursive}
E_j &=- 2\hbar \chi s_0^2 \sqrt{\sum_{k,l} \cos 2 \pi \left(f_k - f_l \right)} \\
&= -2\hbar \chi s_0^2 \sqrt{j +  2 \sum_l \sum_{k<l} \cos 2 \pi  \left(f_k - f_l \right)} \\
&= -2\hbar \chi s_0^2 \sqrt{j}.
\end{align}
The last equality follows from Eq.~\eqref{eqn:energycondition}.  Equation \eqref{eqn:sumofsines} implies that the minimal single-particle energy $E_j$ is achieved when 
\begin{equation}
	f_j = \frac{\delta_j}{2 \pi}.
	\label{eqn:fj}
\end{equation}
The total energy of the chain is $ \sum E_j \propto N^{3/2}$ to leading order in $N$ \cite{Ramanujan1915}. 

\begin{figure}[tb]
	\includegraphics[width=8.6cm]{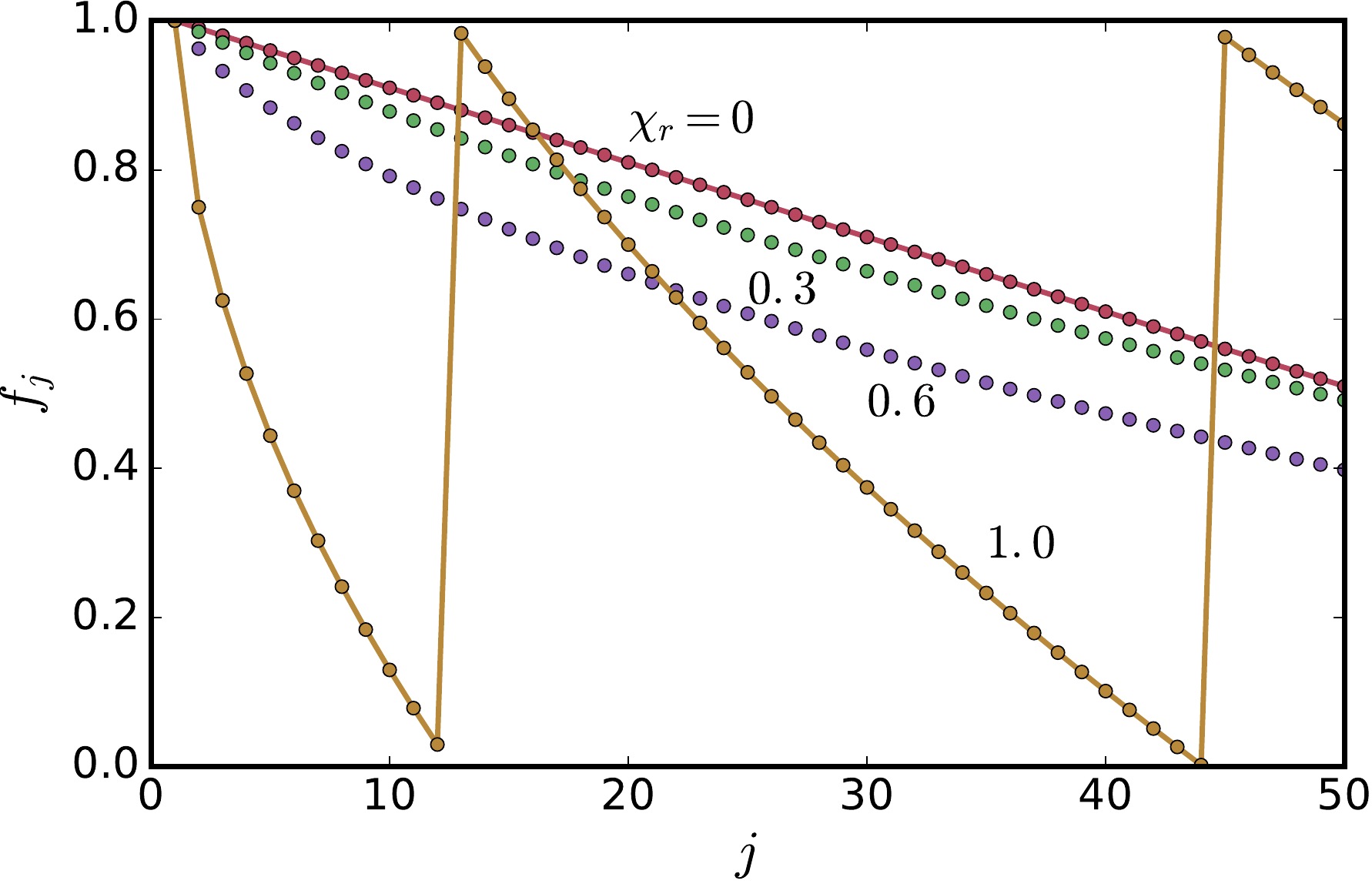}
	\caption{Results of weak-scattering calculations, showing numerical results (points) for a range of chiralities, and analytic results (lines) for $\chi_r = 0, 1$. The $x$-axis enumerates the atoms while the $y$-axis shows their fractional positions as defined in the text.}
	\label{fig:wsresults}
\end{figure}

As shown in \figref{fig:wsresults}, numerical results agree with the positions predicted by Eq.~\eqref{eqn:fj} for both the $\chi_r = 1$ results as calculated by the methods of this section as well as the $\chi_r = 0$ results \cite{Chang2013}. In addition, equilibrium configurations in the weak-scattering limit are included for several intermediate chiralities. For chiralities less than $\chi_r  = 1$, the arrangement of atoms is closer to that of the symmetric case, where the spacing between atoms is uniform and equal to $(1 - \frac{1}{2N}) \lambda $ \cite{Chang2013}. 

The similarity between symmetric and intermediate cases is more pronounced near the right end of the chain, where the spacing between atoms is quite close to uniform. This can be understood intuitively by realizing that atoms on the right side of the chain receive photons from almost all other atoms in both the symmetric and chiral cases, while an atom towards the left receives photons from substantially fewer atoms in the chiral case. Therefore, we might expect atoms further down the chain would resemble the symmetric case more. In addition, it is known that the energy of the symmetric weak-scattering case scales as $N^2$ \cite{Chang2013}, so the fact that it dominates over the chiral configuration (with $E \propto N^{3/2}$) is sensible.
\subsection{Near Resonance}
\label{ssec:nearres}
In the purely chiral case, the system can be solved exactly for any detuning, even outside of the weak-scattering limit discussed in Sec.~\ref{sec:ws}. Consider the equations of motion for a single atom $j$ in the purely chiral case:
\begin{equation}
\dot{p}_j =   - \hbar k \chi \abs{\sge{j}}^2 -2 \hbar k \chi  \Re \sum_{i<j} \seg{j} \sge{i} e^{i k \left(z_j - z_i \right)} - \gamma_p p_j,
\end{equation}
\begin{equation}
\dot{\sigma}^j_{ge} = \left( i \delta - \frac{\gamtot}{2} \right) \sge{j} + i \Omega - \chi \sum_{i < j} \sge{i} e^{i k \abs{z_j - z_i}} .
\end{equation}
Setting both time derivatives to zero gives
\begin{equation}
p_j = - \frac{ \hbar k \chi}{\gamma_p} \left[ \abs{\sge{j}}^2 +2 \Re \sum_{i<j} \seg{j} \sge{i} e^{i k \left(z_j - z_i \right)} \right],
\label{eqn:purechimomentum}
\end{equation}
\begin{equation}
\sge{j} = \frac{ - i \Omega + \chi \sum_{i < j} \sge{i} e^{i k \abs{z_j - z_i}} }{i \delta - \frac{\gamtot}{2}}.
\label{eqn:sgej}
\end{equation}
Once again, all the $p_j$'s must be the same. Therefore all $p_j$ = $p_0$, and since the equations of motion for $p_0$ do not refer to the position of the other atoms, it can be found to be
\begin{equation}
p_j = - \frac{\hbar k \chi}{\gamma_p} \abs{\sge{0}}^2,
\end{equation}
where the value of $\sge{0}$ follows from the $j = 0$ case of Eq.~\eqref{eqn:sgej}.
By substituting this into Eq.~\eqref{eqn:purechimomentum}, the general condition for all atoms is found,
\begin{equation}
\abs{\sge{j}}^2 - \abs{\sge{0}}^2 = -2 \hbar k \chi  \Re \sum_{i<j} \seg{j} \sge{i} e^{i k \left(z_j - z_i \right)}.
\end{equation}

The damping rate $\gamma_p$ dropped out of the equations of motion entirely, just as in the weak-scattering case. 
It is also important that the equations of motion for $j$ do not make any references to an atom $i > j$. 
Thus these equations can be  easily solved iteratively starting from the first atom, $j = 0$, for any value of $\delta$.
\begin{figure}[tb]
	\includegraphics[width=8.6cm]{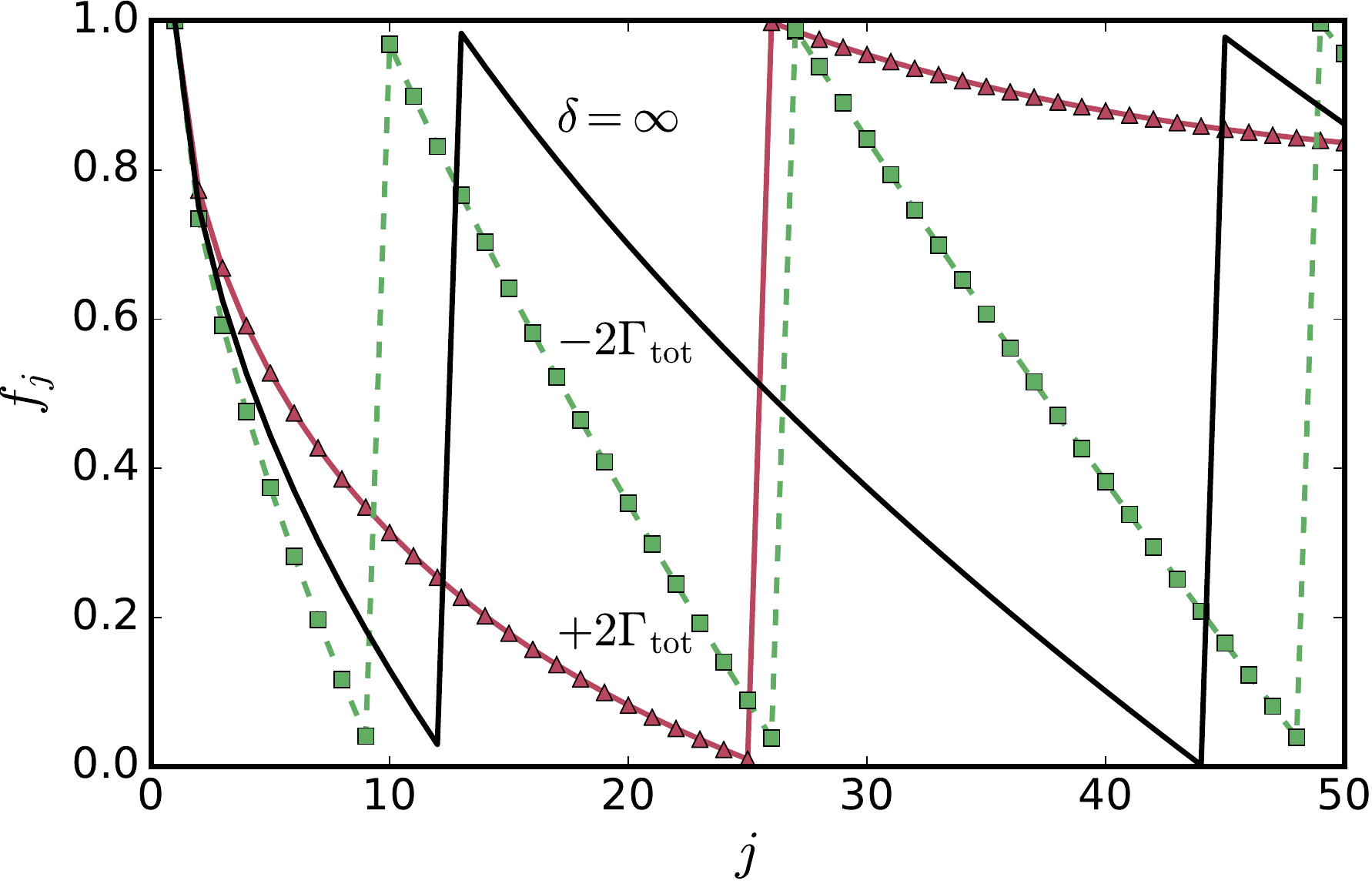}
	\caption{Results of calculations near-resonance at full chirality ($\chi_r = 1$), showing numerical results (points) and analytic results (lines). The weak-scattering case $\left( \delta \to \infty \right)$ has been included for comparison.}
	\label{fig:alldetunings}
\end{figure}
This iterative approach requires a purely one-way coupling $\left( \chi_r = 1 \right)$.
As shown in \figref{fig:alldetunings}, this analytical result agrees with the numerical results precisely, although closer to resonance the numerical method does not converge to a self-organized solution.

\begin{figure}[tb]
	\includegraphics[width=8.6cm]{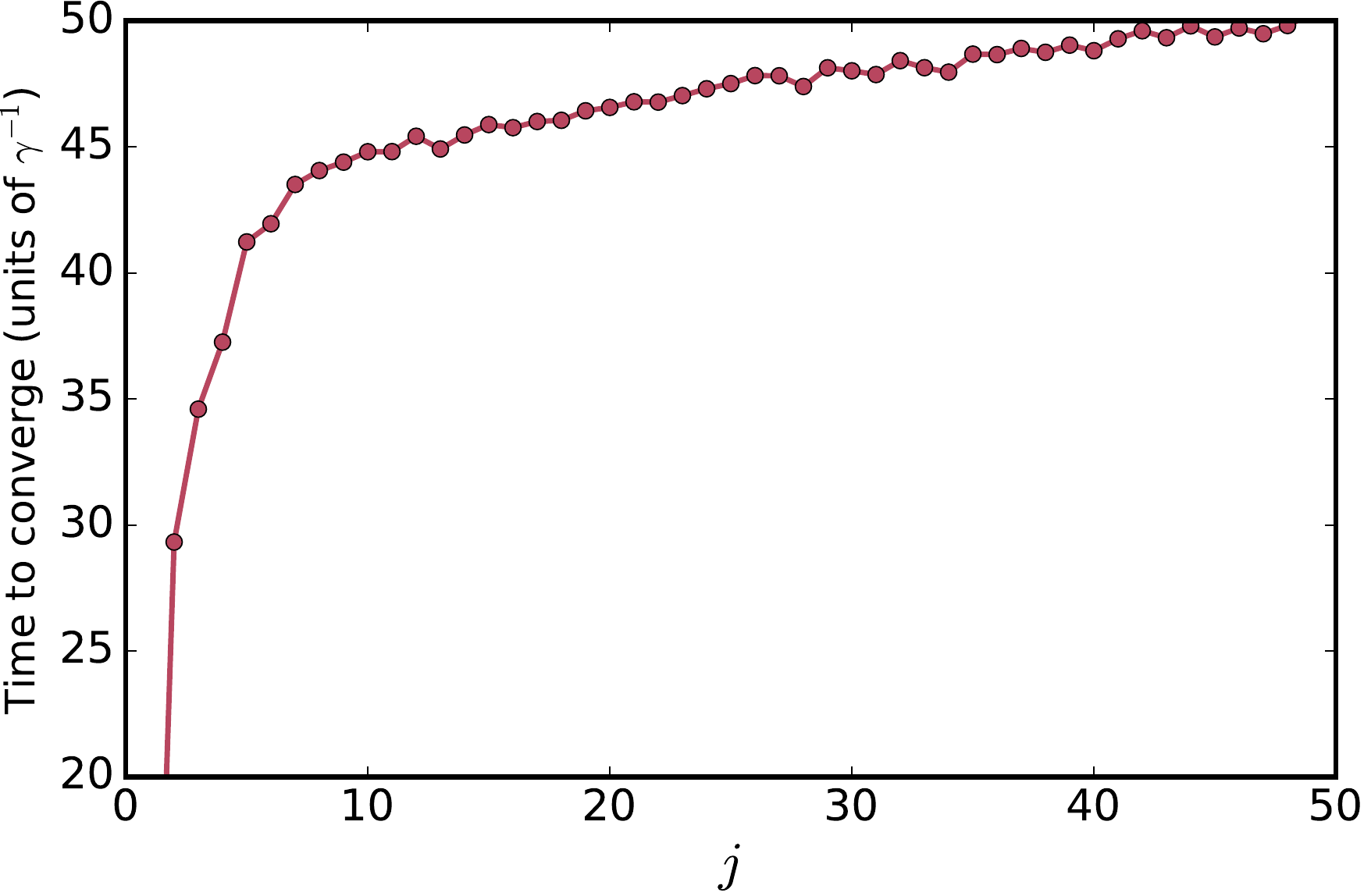}
	\caption{The time required to converge to equilibrium for each atom. This time is defined as the latest time at which the atomic momentum is more than 1\% away from its equilibrium value. This data represents 10,000 separate instances of the weak-scattering limit with $\chi_r = 1$ in random initial positions each time. For each trial, the time for the first atom to converge to its final position was calculated, and this was then subtracted from all data points to isolate the trends. Atoms further down the chain converge later in the process.}
	\label{fig:ttc}
\end{figure}

The iterative nature of the calculation appears to have physical significance. In simulation, each atom tends to come to equilibrium ``one by one,'' with an atom equilibrating later if further down the chain. This trend in equilibration times is illustrated for the weak-scattering case in \figref{fig:ttc}.

\section{Small Chiralities Near Resonance}
\label{sec:arbitrarychiralities}
In Sec.~\ref{sec:ws}, we discuss results at arbitrary chirality for pump fields far from resonance, while Sec.~\ref{ssec:nearres} focused on cases near resonance only for fully chiral ($\chi_r = 1$) cases. In this section, we turn our attention to the general case, with pump fields at arbitrary detunings and $\chi_r$ any value.

For $\chi_r \gtrsim .65$ the system does not converge to a steady-state in the weak-scattering regime. 
Instead, a limit cycle behavior is observed, in which the atomic configuration periodically returns to previous points, but it does not settle into a steady-state or self-organized behavior. Similar limit cycle behavior in driven dissipative systems was observed in Ref.~\cite{Chan2015}. 
In general, the limit cycle behavior emerges at $\chi_r > .6$ and vanishes at $\chi_r > .99$. The locations of these dynamical transitions are sensitive to the number of atoms used in the simulation. This behavior makes our protocol for starting in a far-detuned pump field and slowly stepping down closer to resonance unviable, and we do not examine such chiralities further.  

For smaller chiralities at most detunings, there are not large differences between cases with intermediate chirality $\chi_r < 0.65$ and the symmetric case. Much of the relevant behavior and analysis is contained in Ref.~\cite{Chang2013}.

\begin{figure}[tb]
	\includegraphics[width=8.6cm]{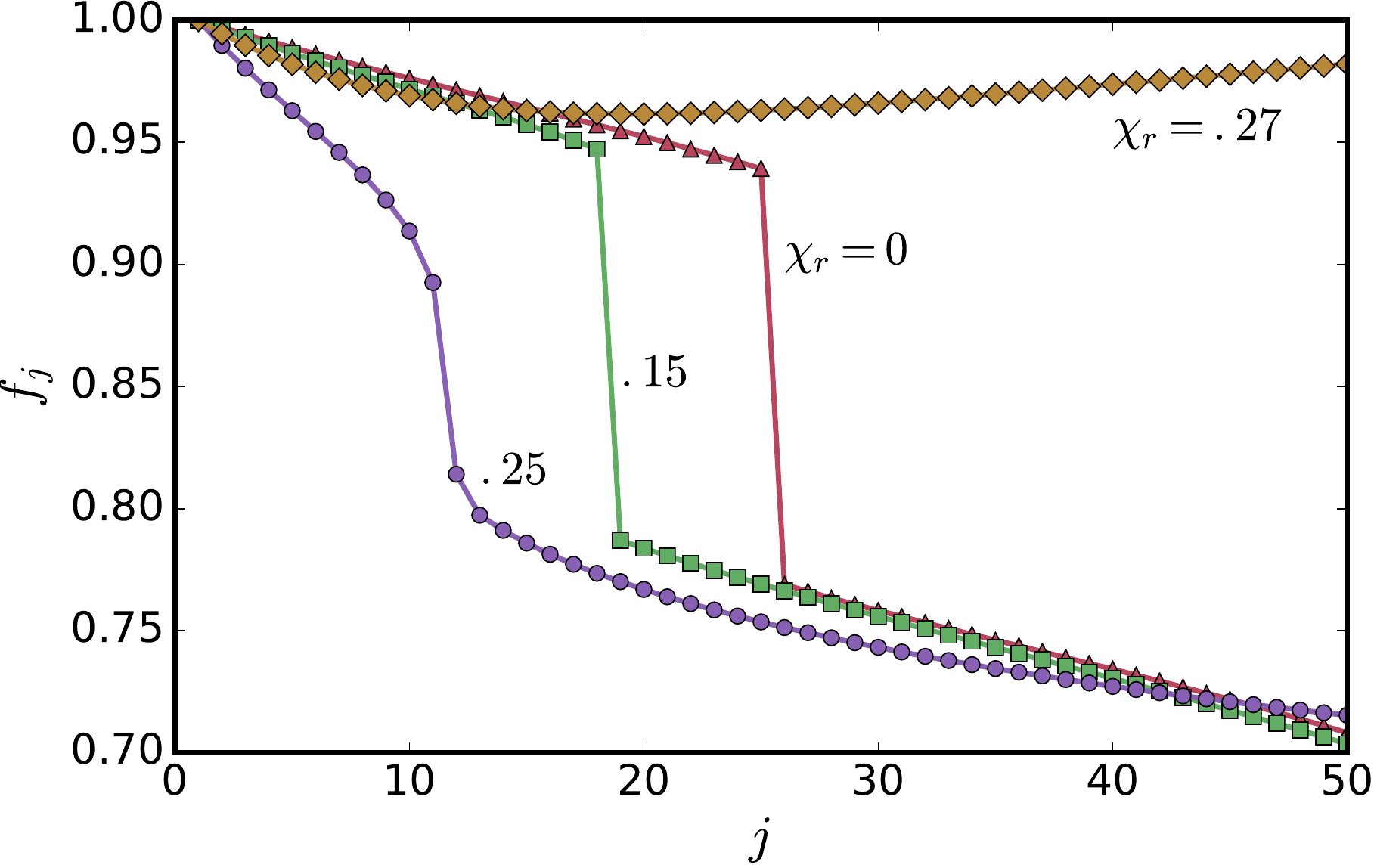}
	\caption{Evolution of the phase-slip configurations as $\chi_r$ increases from 0 to $0.27$ at $\delta = \gamtot$. These are produced by the numerical protocol outlined in Section \ref{sec:numerics}.}
	\label{fig:chiralsa}
\end{figure}

For small, positive detunings, small changes in chirality can produce drastic differences in behavior. In the case of a symmetric interaction, the atoms form a ``phase-slip'' configuration in which two halves of the atomic ensemble form a nearly regular lattice with spacing $\lambda$, with a $3 \lambda/4$ phase slip between them. If the two collections of atoms were single bodies interacting with a dipole-dipole potential, then their interaction would be proportional to $\sin 2 \pi (f_2 - f_1)$, and the $3 \lambda / 4$ positional difference between them minimizes this two-body potential. The phase-slip configuration does not occur at large chiralities, but can emerge at lower nonzero chiralities. As chirality increases from $\chi_r = 0$, an $N$ atom chain will no longer form two $N/2$ groups, but that instead the groups become increasingly imbalanced until the phase slip becomes unstable at $\chi_r \approx 0.3$ and collapses to a single lattice with spacing close to $\lambda$ (\figref{fig:chiralsa}).

The configurations in \figref{fig:chiralsa} are obtained by beginning far off resonance and slowly decreasing $\delta$ as described in \ref{sec:numerics}. However, we can also converge first to a phase-slip configuration close to the critical chirality (where the phase-slip configuration disappears) and then use this as the initial condition for a simulation with slightly higher chirality. This will still lead to a collapse at roughly the same $\chi_r$, but it allows the collapse process to be observed. Collapse due to a quench in chirality from $\chi_r = 0.26$ to $\chi_r = 0.27$ is a runaway process, in which atoms cross the phase slip one by one. This process cascades until there is only one chain and no phase slip. 

\begin{figure}[tb]
	\def\stackalignment{l}
	\topinset{\includegraphics[width=3cm]{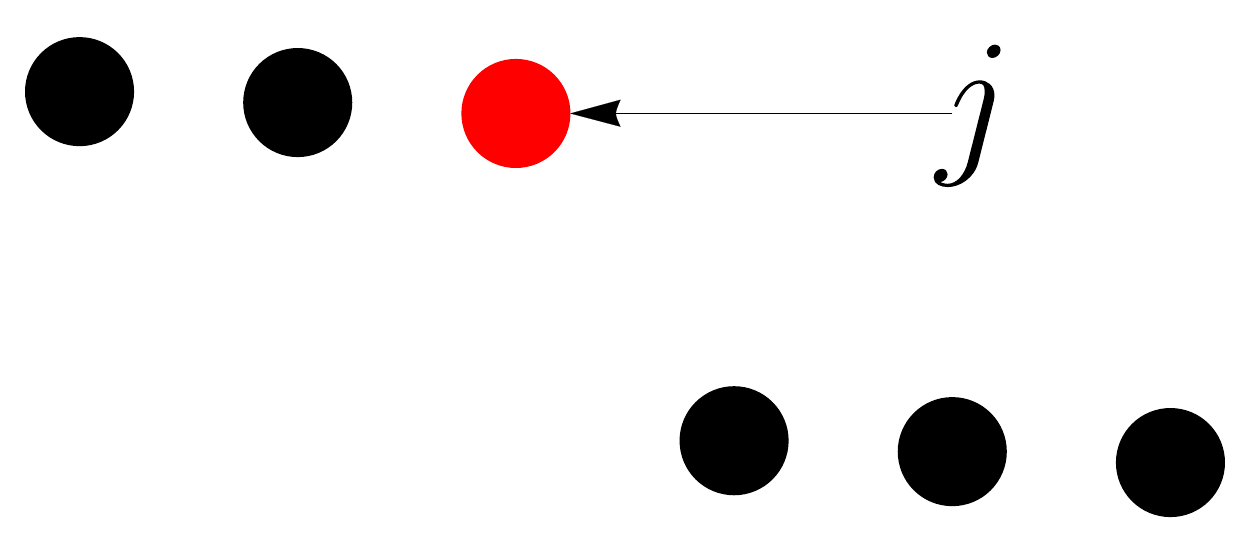}}{\includegraphics[width=8.6cm]{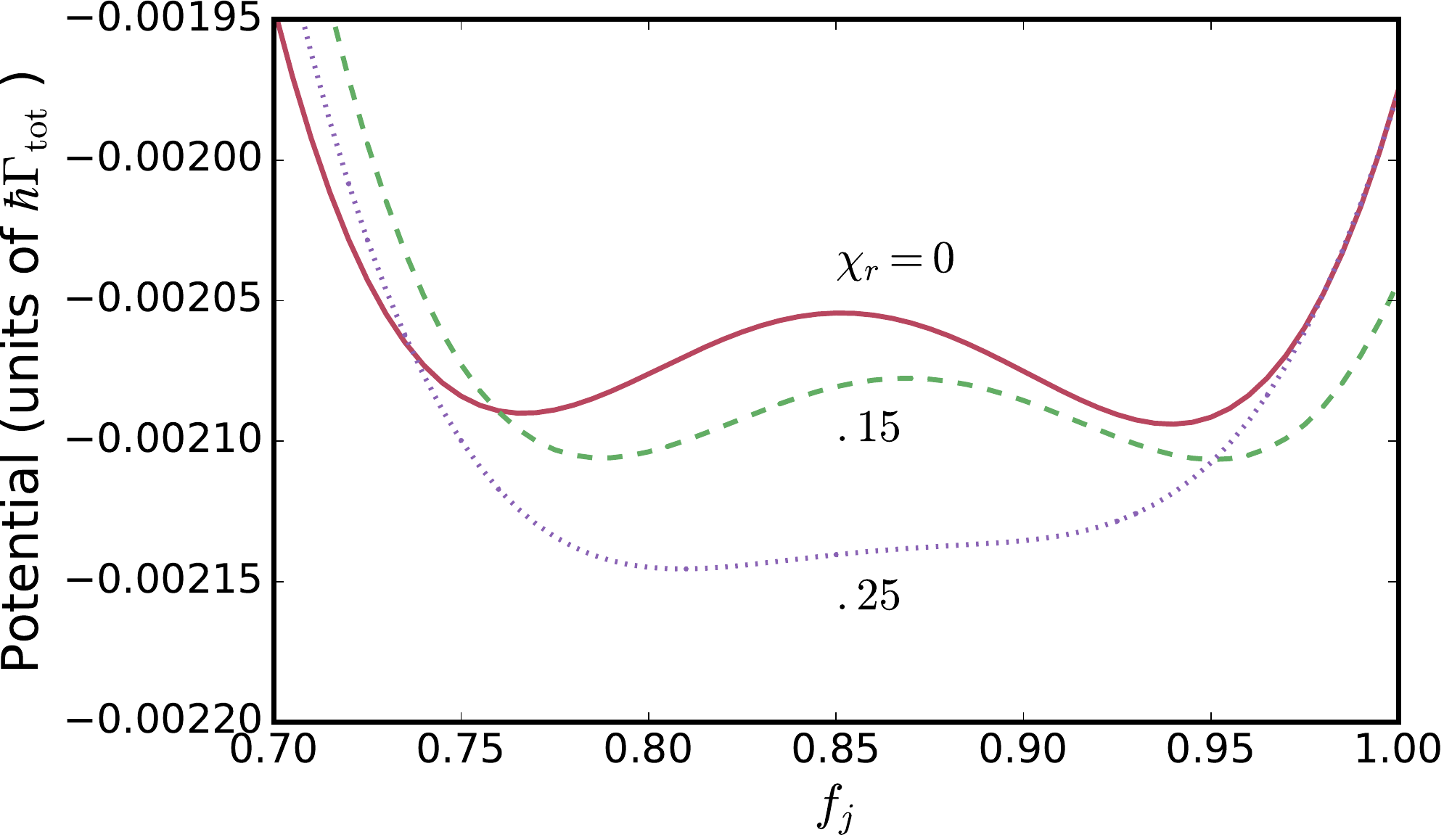}}{4pt}{1.75cm}
	\caption{The effective potential resulting from integrating the force equation Eq.~\eqref{eqn:eomp} in various equilibrium configurations, ranging from symmetric ($\chi_r = 0$) to just before collapse ($\chi_r = 0.25$) at detuning $\delta = \gamtot$. This curve was generated by examining the force on the last atom before the phase-slip (the rightmost atom in the left super-atom), as illustrated in the inset. Note that a constant vertical offset on any of these lines has no physical significance.}
	\label{fig:wells}
\end{figure}

A heuristic picture offers insight into the process of this collapse. Consider the equation of motion for atom $j$. If other atoms are frozen in their equilibrium positions and coherences, the force atom $j$ would experience if it were moved away from its original $z_j$ can be determined by calculating $\dot{p}_j$. (Note that we the equilibrium coherence at each $z_j$ should also be recalculated.) By integrating the resulting curve, an effective potential is obtained. The problem is \textit{not} actually well-described by a potential, and this effective potential includes nonconservative terms like the damping force. This process can be performed for any atom $j$, but targeting the one just before the phase slip allows us examine its stability against small perturbations.
At $\chi_r = 0$, the equilibrium configuration starts out with a ``double well'' potential (\figref{fig:wells}), with both sides stable. As chirality increases, one well weakens and eventually vanishes leaving only a single stable point.
The behavior of the effective potential should be compared to the number of stable configurations at each point in parameter space. For $\chi_r = 0$, the $N/2$ split is not the only stable configuration; as many as seven atoms out of a total of fifty can be moved from one side to the other and still maintain a stable phase-slip configuration. At the other extreme, at $\chi_r = 0.25$, there are only two stable phase-slip configurations. This movement of atoms across the phase slip is accomplished by converging to a stable solution at a particular $\chi_r$, and then manually setting $f_j = f_{j+1}$, moving the atom across the phase slip. This modified steady state is then used as the initial condition for another run of the simulation to ensure that another steady state exists with this configuration. Situations that exhibit a double-well structure allow the atom to be placed on either side of the phase-slip, but this freedom slowly disappears as $\chi_r$ increases.

\section{Experimental Considerations}
\label{sec:experimentalconsiderations}
In this section, we first outline a scheme by which tunable chirality can be achieved in the optical nanofiber platform (Sec.~\ref{ssec:varchi}). We also present an analysis of the optical response of self-organized ensembles by transfer matrix methods (Sec.~\ref{ssec:tfermat}) and consider the motional modes in the weak scattering regime (Sec.~\ref{ssec:motion}). Finally, we consider other experimental concerns including the influence of finite temperature (Sec.~\ref{ssec:other}). Throughout this section, we use the $\mathrm{D}_2$ line of neutral cesium, with details given in Ref.~\cite{Steck2010b}.
\subsection{Achieving Variable Chirality}
\label{ssec:varchi}

\begin{figure}[tb]
	\centering
	\includegraphics[width=6.0cm]{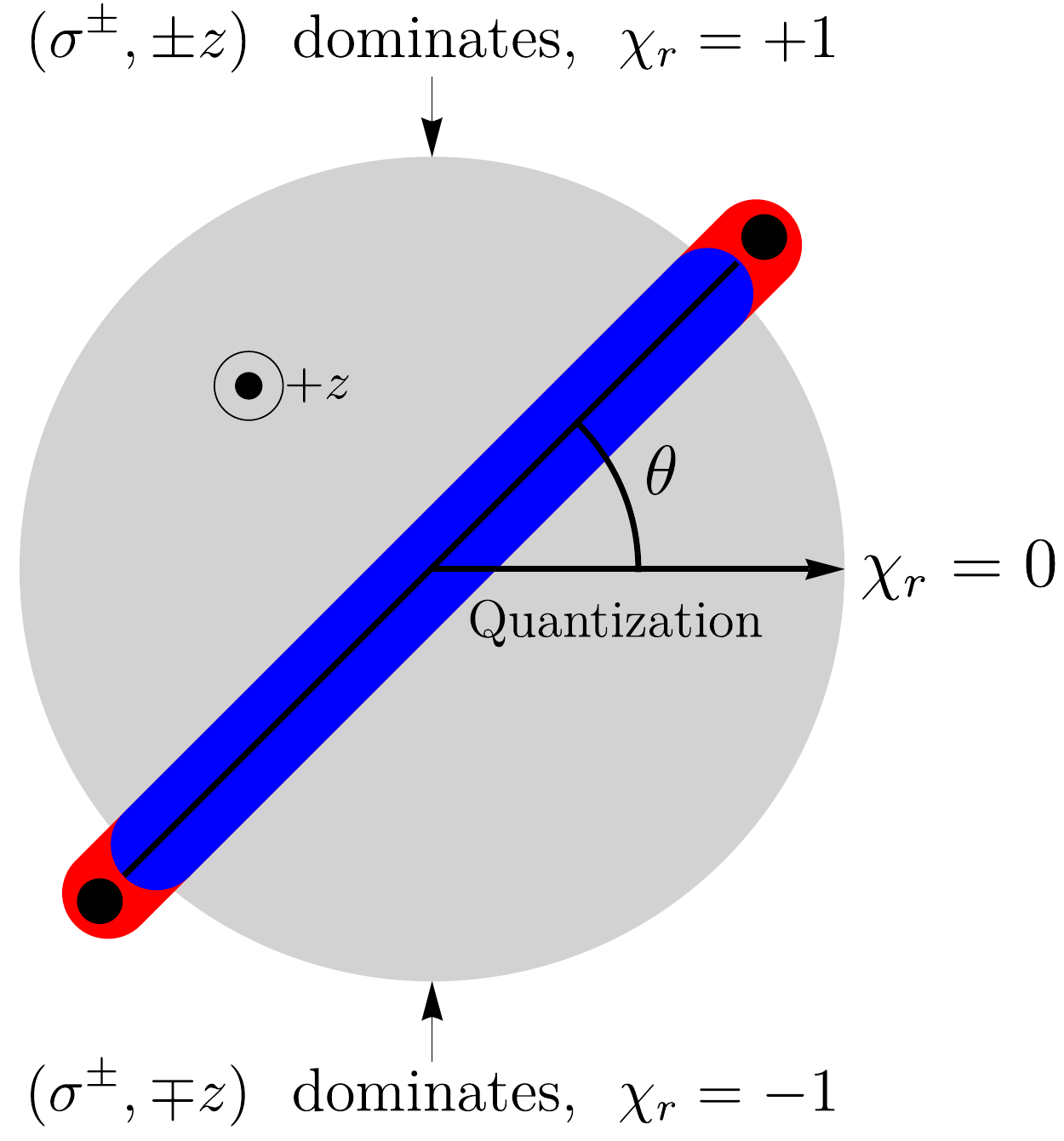}
	\caption{Illustration of how a variable polarization of trapping light can be used to achieve the full range of chiralities. The central grey circle shows a cross-section of the nanofiber. The elongated red-blue shape represents the trapping potential, which can be rotated by choosing a quasilinear polarization of the trapping light rotated by $\theta$ relative to the quantization axis. Since sweeping through this angle varies the coupling from one chirality to the next continuously, the value of $\chi_r$ will vary from $-1$ to $1$. The positive $z$ axis points out of the page.}
	\label{fig:varchirality}
\end{figure}

In this work, $\chi_r$ is treated as a tunable chirality, but we have not discussed how this might be achieved. In many systems, this can be achieved by taking advantage of spin-orbit coupling in light. A thorough theoretical treatment of this phenomenon can be found in Refs.~\cite{Bliokh2015,Bliokh2014}. Experimental measurement of chiral behavior in nanofibers is described in Ref.~\cite{Mitsch2014a}, while similar measurement controlling directional emission from a nanobeacon was performed in Refs.~\cite{Leuchs2014,Neugebauer2014}. An external magnetic field breaks cylindrical symmetry and provides a quantization axis (horizontal in \figref{fig:varchirality}) by setting the direction of the two-level atomic dipole. Due to the confinement of light in the fiber, its polarization has a longitudinal component which is parallel to the wavevector $\vec{k}$ (in or out of the page in \figref{fig:varchirality}) and a transverse component perpendicular to the quantization axis. If the longitudinal and transverse amplitudes are equal (as shown at the top of \figref{fig:varchirality}), this yields a circular polarization relative to the quantization axis. However, if the direction of $\vec{k}$ is reversed then the two components (transverse and longitudinal) will now change phase with respect to each other. Therefore, light propagating in opposite directions will have opposite circular polarizations (opposite signs in text on top of \figref{fig:varchirality}). Similarly, at a diametrically opposite position on the fiber surface (bottom of \figref{fig:varchirality}), the component of polarization transverse to $\vec{k}$ has been reversed. Thus, at the bottom of \figref{fig:varchirality} the pairing of circular polarization and direction of propagation will be reversed compared to the top of the figure. At angular positions between these two positions, the two modes are mixed, meaning light traveling in either direction will consist of both polarizations. An atom can be coupled to a given light field with variable, controllable chirality by adjusting its azimuthal position.

Consider using a cesium atom in its electronic ground state, $6^2 S_{1/2}$. Under a pump field resonant to the $\ket{ 6 ^2S_{1/2} , F = 4, m_F = 4} \to \ket{ 6 ^2 P_{3/2} ,F = 5, m_F = 5}$ transition, excited electrons can only decay into the $\ket{6^2 S_{1/2},F = 4, m_F = 4}$ state. In addition, this decay can only occur via the emission of a  $\sigma^+$ polarized photon. If an atom is trapped on one side of the fiber (the ``top'' side in \figref{fig:varchirality}), then it predominantly emits in the $-z$ direction. On the opposite side atoms predominantly emit in $+z$. At an intermediate position (i.e., along the quantization axis), atoms have access to both modes equally, and decay symmetrically. Adjusting the azimuthal position of the atomic ensemble relative to the quantization axis varies the coupling from symmetric to chiral. This is possible by adjusting the polarization of the trapping light, which entirely controls the azimuthal position of the trap. This scheme is similar to that used by previous experimental realizations of chirality \cite{Young2014,Petersen2014}.

\subsection{Optical Signatures}
\label{ssec:tfermat}
Next, we establish a method by which chiral self-organization can be detected through its optical response to a weak probe field. Ordered spatial configurations of atoms can be probed by sending weak guided light through the ensemble and looking at changes in reflection and transmission compared to both a disordered state and a different ordered configuration \cite{Corzo2016,Qi2016,Sørensen2016}. The optical properties of the atomic ensemble can be calculated using a transfer matrix formalism \cite{Deutsch1995,Asboth2008}. It is natural to assume that the atomic response to the probe field will have the same chirality as the light responsible for self-organization, which requires transfer matrix equations incorporating a chirality \cite{Corzo2016,LeKien2014a}. However, if the probe has the same chiral properties as the light used for pumping, the interaction with probe light changes with chirality. There are then two variables changing with $\chi_r$: the spatial configuration of the atoms in equilibrium and the nature of their interaction with the probe. Therefore, for what follows, we set $\Gamma_L = \Gamma_R = \Gamma$ for the optical response to the probe light. Physically, the probe light might differ in either polarization or mode structure from the light which mediated self-organization, in such a way that the interaction it has with the ensemble is not chiral. This allows us to see the influence only of the change in equilibrium configuration. 

\begin{figure}[tb]
	\includegraphics[width=6.0cm]{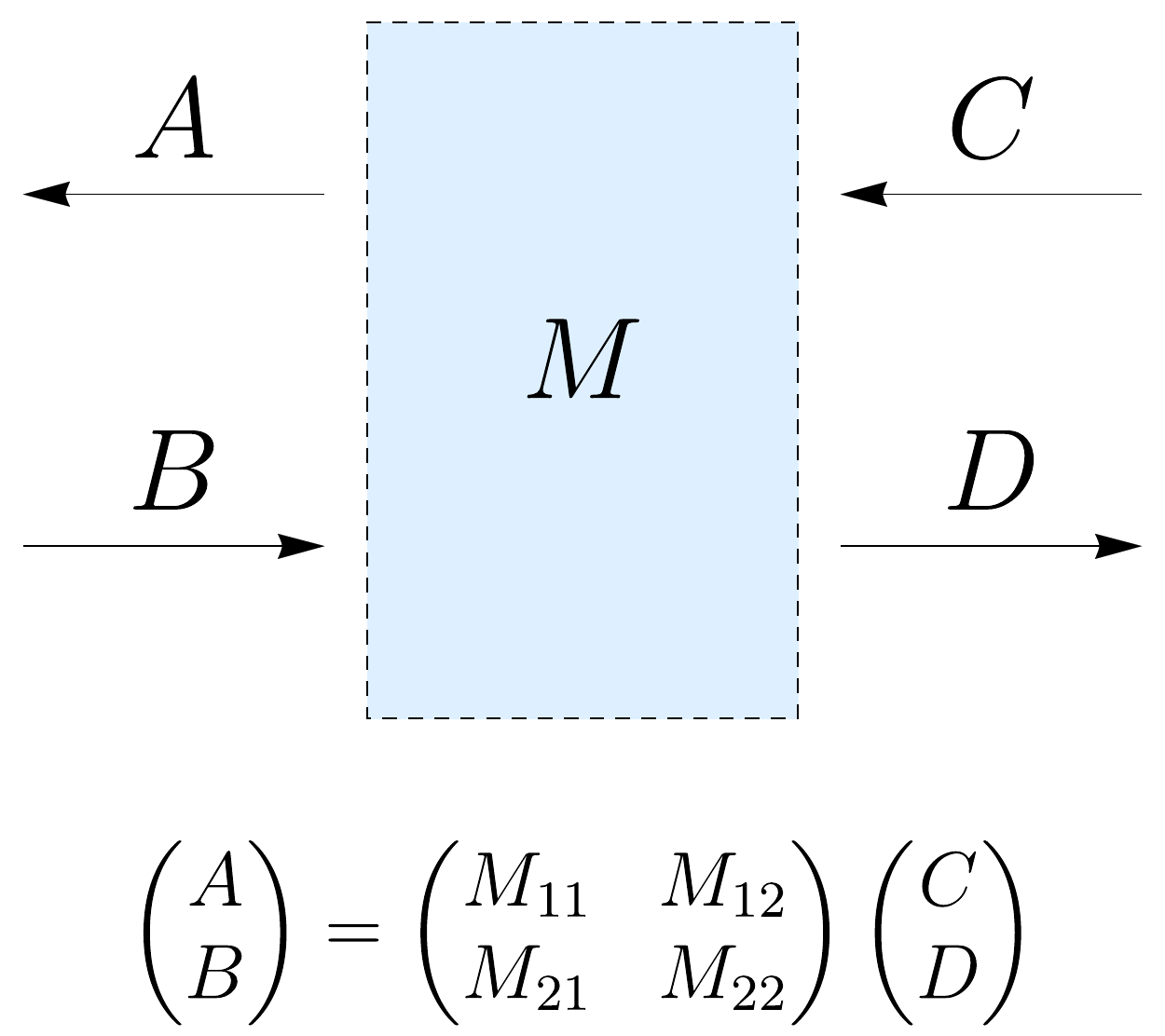}
	\caption{An illustration of how the transfer matrix can be used to calculate the fields on one side of an optical element given fields on the other side.}
	\label{fig:tfermat}
\end{figure}

Assuming such symmetric probing, the transfer matrix for a single atom in response to a probe field of detuning $\Delta$ is given by
\begin{equation}
	m = \frac{1}{t} \begin{pmatrix} t^2 - r^2 & r \\ - r & 1 \end{pmatrix}.
\end{equation}
Here, the transmission and reflection coefficients are
\begin{equation}
	r = - \frac{2 \Gamma}{2\Gamma + \gamma - 2 i \Delta}
\end{equation}
and
\begin{equation}
	t = 1 - \frac{2 \Gamma}{2 \Gamma + \gamma - 2 i \Delta}.
\end{equation}
This transfer matrix computes fields on the left of an optical element using the value of fields on the right, as shown in \figref{fig:tfermat}. 

The transfer matrix for an entire ensemble is the product of these individual atomic transfer matrices as well as others corresponding to free propagation over a distance $d$, which accounts for accumulated phase between atoms:
\begin{equation}
	\begin{pmatrix}
		e^{i k d} & 0 \\ 0 & e^{-ikd} \end{pmatrix}.
\end{equation}
This calculation yields a transfer matrix $M$ characterizing the optical response of the entire atomic array. From $M$ the optical properties of the ensemble can be extracted by making an analogy to the single-atom transfer matrix. If the elements of the final matrix are $M_{ij}$, the derivation for the original single-atom transfer matrix can be reversed to find the transmission $T$ and reflection $R$ coefficients for the calculated transfer matrix in terms of those elements \cite{LeKien2014a,Corzo2016}:
\begin{align}
	T_R &= \frac{1}{M_{22}}, \\
	R_L &= -\frac{M_{21}}{M_{22}}, \\
	R_R &= \frac{M_{12}}{M_{22}}, \\
	T_L &= M_{11} - \frac{M_{12} M_{21}}{M_{22}}.
\end{align}
Here, $R$ and $L$ refer to the transmission and reflection response to right- or left- propagating light. Due to the atomic arrangement, the response to light on either side of the optical element formed by the atomic ensemble may not be symmetric, even if the individual atomic responses are.

\begin{figure}[t]
	\includegraphics[width=8.6cm]{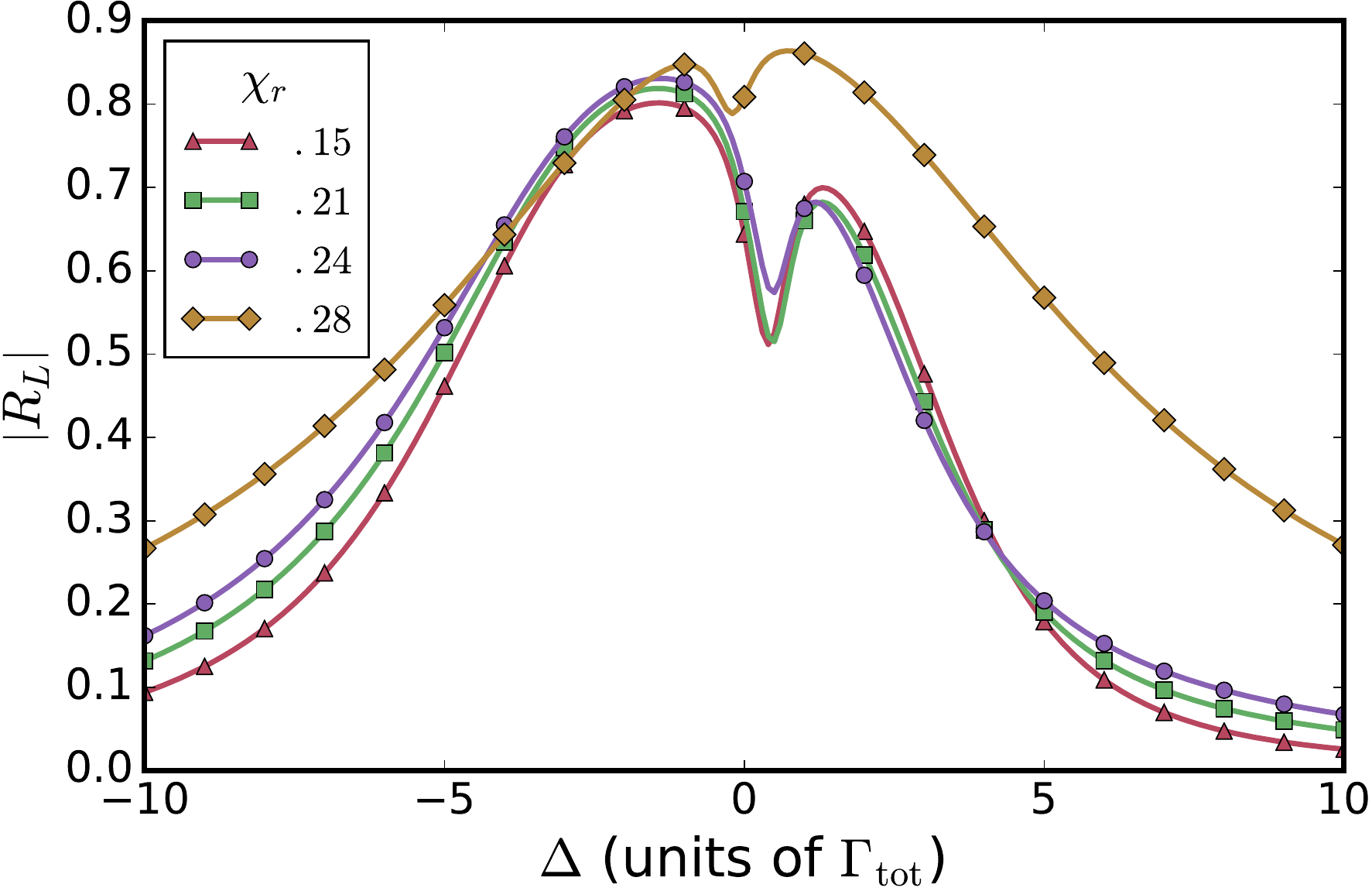}
	\caption{Reflection spectra for configurations near ($\chi_r = 0.15, 0.21, 0.24$) and just past $(\chi_r = 0.28)$ the super-atom collapse ($\delta = \gamtot$), calculated via transfer matrix technique. While these plots are quite similar for both directions, the quantity plotted here is $\abs{R_L}$.}
	\label{fig:satransfer}
\end{figure}

The collapse of phase-slip configurations as $\chi_r$ increases is reflected in optical response by a sudden increase in the width of the reflection peak, as shown in \figref{fig:satransfer}. This can be understood by the well-known result that evenly spaced atoms form a mirror due to Bragg reflection \cite{Deutsch1995,Birkl1995,Chang2012}. The quality of this mirror grows as the number of atoms in the ensemble increases. After the collapse, two small mirrors with a phase slip between them become a single larger mirror, with a corresponding increase in the reflective effectiveness.

\begin{figure}[b]
	\includegraphics[width=8.6cm]{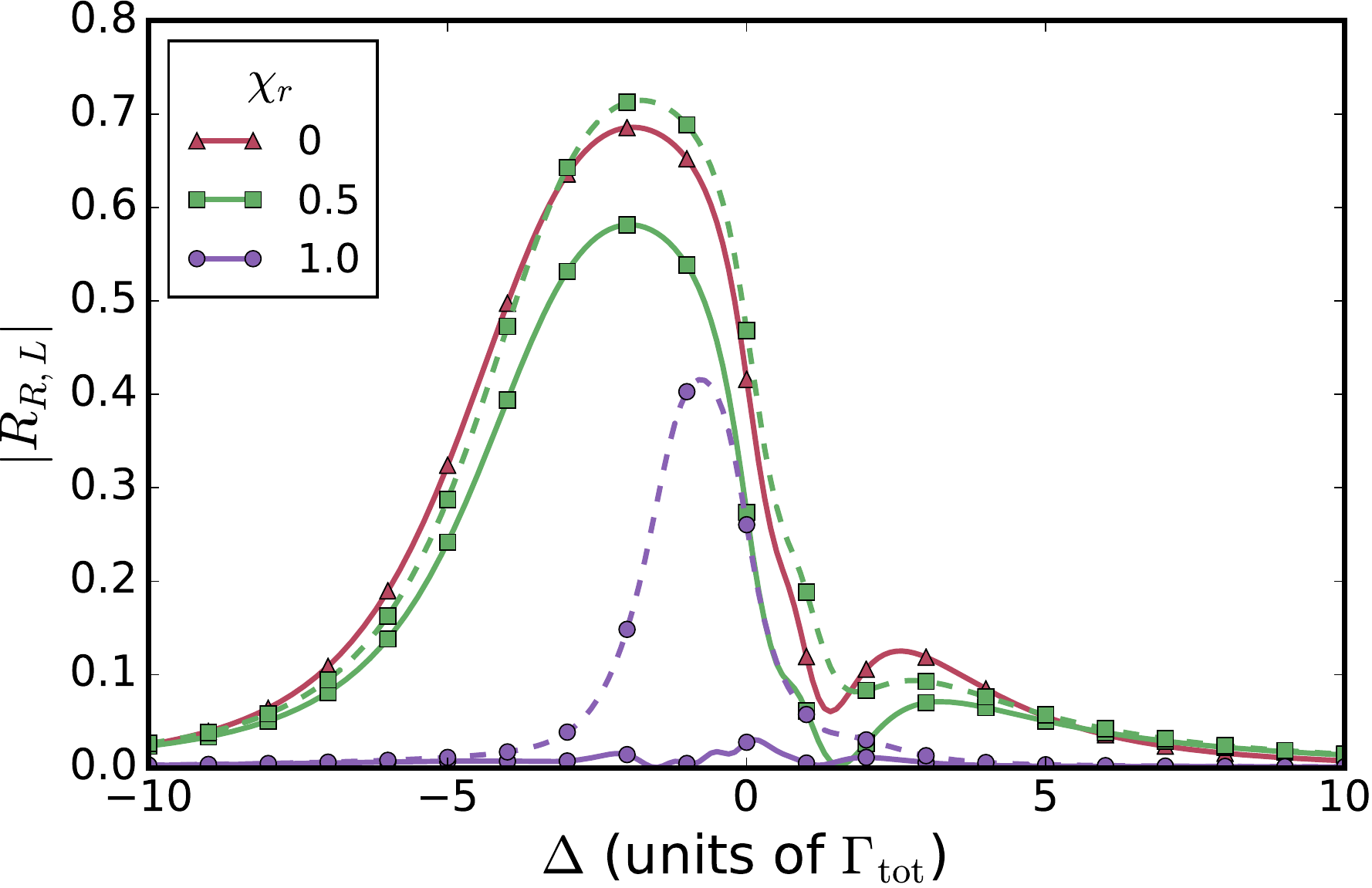}
	\caption{Reflection spectra for weak-scattering configurations, calculated via transfer matrix. Dashed lines represent $\abs{R_L}$, the reflected amplitude of left-propagating light, while solid lines represent $\abs{R_R}$.}
	\label{fig:wstransfer}
\end{figure}

Results from the weak-scattering regime are shown in \figref{fig:wstransfer}. Even assuming a symmetric atomic response to probe light, the ensemble responds differently to different directions of incoming light when it has been self-organized under highly-chiral (e.g., $\chi_r = 1$) interactions. One way to understand this asymmetry is to consider the behavior of the light as it impinges on the $\chi_r = 1$ arrangement shown in \figref{fig:wsresults}. Because the average atomic spacing approaches an evenly spaced lattice on one side of the ensemble, light that encounters that side first is primarily Bragg reflected. From the other direction, however, the light first encounters an irregularly spaced lattice of atoms which scatter the light into free space.

If, unlike the previous cases, the probe and the atomic coupling are both chiral, then the probe light will be strongly scattered when incident in the direction that atoms are sensitive to. At the same time, light passing in the opposite direction will not interact with the atoms. In this case, light can pass in one direction but not the other. The ability of chiral systems to provide a direction-dependent optical element has been demonstrated before \cite{Sayrin2015}, but not in a self-organized context.
\subsection{Vibrational Modes}
\label{ssec:motion}
Vibrational modes also allow a trapped atomic ensemble to be interrogated, most obviously by the appearance of motional sidebands in the atomic spectrum. Also, an atom driven at twice a trap frequency will undergo parametric heating \cite{Jauregui2001,Savard1997}. Because this resonant effect is sensitive to the trap frequencies, it can be used for measurement and characterization of trap parameters \cite{Friebel1998,Roati2001}. Since the atoms in a self-organized configuration are trapped by a potential generated by the other atoms, parametric heating provides information about the organizing potential. These modes may also have use in quantum information processing or storage \cite{Muschik2011}. Therefore, in this section, we examine the structure of the phonon spectrum for atoms in the chiral case. 

A full treatment by matrix diagonalization of the modes in a symmetric interaction can be found in Ref.~\cite{Chang2013}. For the fully chiral case in the weak-scattering limit, simple expressions for all the motional modes exist. Without a damping term, the first term in Eq.~\eqref{eqn:eomp} produces a constant acceleration which is identical for all atoms. Since this change in reference frame does not affect the motional modes, we drop this term and impose the weak-scattering limit, reducing Eq.~\eqref{eqn:eomp} to
\begin{equation}
	\dot{p}_j = - 2 \hbar k \chi s_0^2 \sum_{i < j} \cos k (z_j - z_i).
\end{equation}
This force is equivalent to atom $j$ experiencing the potential given in Eq.~\eqref{eqn:sumofsines}. Expanding that potential to second order around equilibrium yields a natural frequency of oscillation:
\begin{equation}
	\omega_j = \sqrt{4 \chi s_0^2 \omega_r \sqrt{j}},
\end{equation}
where $\omega_r = \hbar k^2 / \left(2 m\right)$ is the recoil frequency of the atom. 

Only one $\omega_j$ directly emerges from the potential formulation. However, atoms to the left of $j$ will oscillate  at other frequencies $\omega_i$, modulating the equilibrium position of $z_j$. Therefore, atomic motion from one atom also contributes to the motion of atoms further to the right, and the motion of $z_j$ will contain frequency components $\omega_i$ for all $i < j$.

\begin{figure}[tb]
	\includegraphics[width=8.6cm]{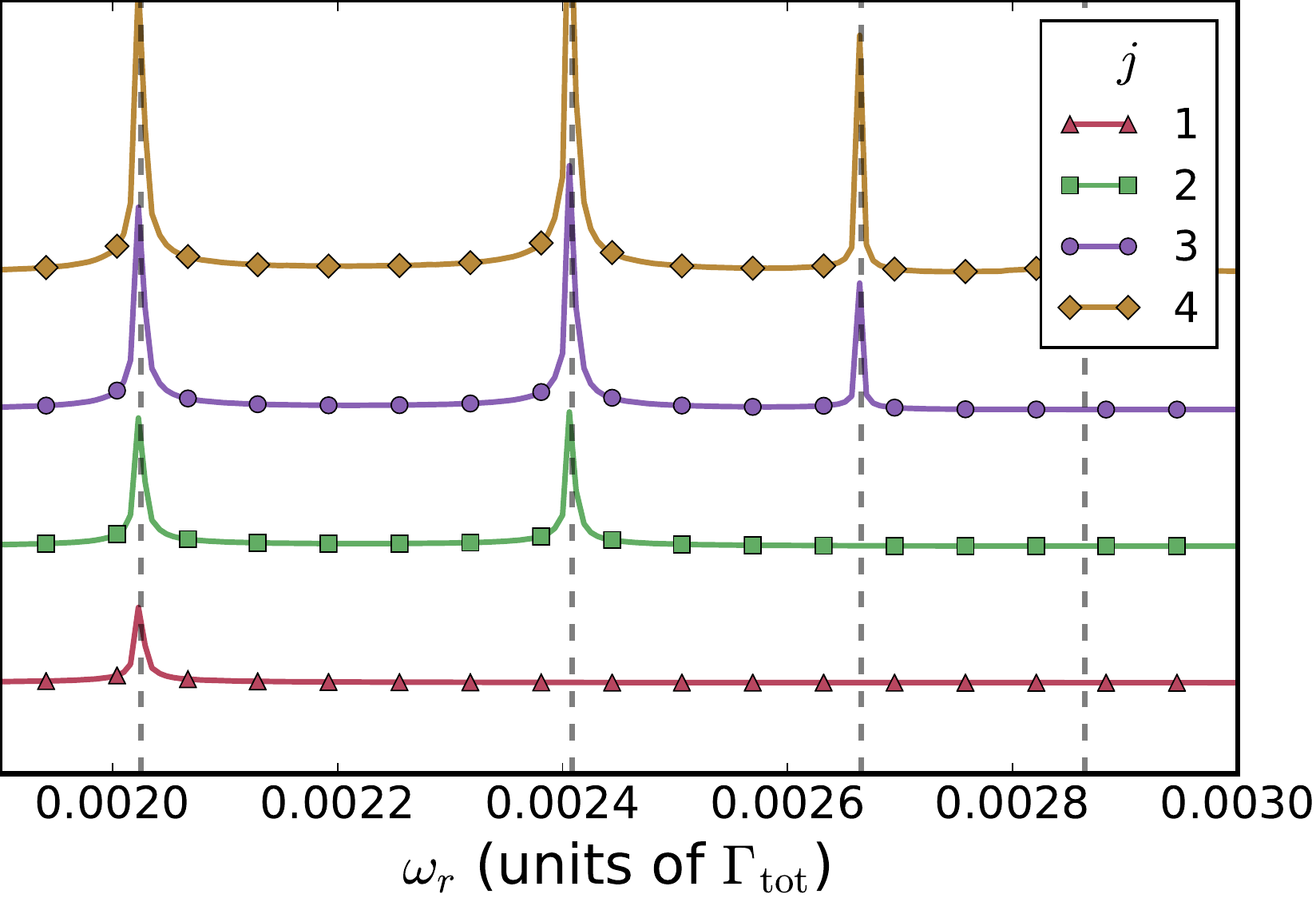}
	\caption{Fourier transform (arbitrary units) of the motion of the leftmost five atoms in a fully chiral chain ($\chi_r = 1$) in the weak-scattering limit. Atoms were brought to equilibrium positions, damping was then turned off, and they were displaced slightly from equilibrium. Vertical dashed lines represent theoretical values where frequencies are expected, while numbers index both the frequencies and (by color) denote which line corresponds to which atom's motion. Motion of the leftmost atom (j = 0) is not visible because it undergoes only uniform acceleration and is thus used as a reference point. The curves are shifted vertically relative to each other for ease of visualization.}
	\label{fig:motion}
\end{figure}

These multiple frequencies are precisely the behavior seen if every atom is perturbed slightly from equilibrium (\figref{fig:motion}). The second atom has only one frequency present and the others have increasingly more. Note that the motion of the leftmost atom is not visible, as we set $f_0 = 1$ when defining coordinates. Since it experiences no potential, it has no natural frequency of oscillation. Thus, experimental procedures addressing these sidebands would be specific to a subset of atoms. Unlike the symmetric case, in which the frequency of the motional modes is proportional to $\sqrt{N}$ for large $N$, the chiral case has frequencies which are not changed by the addition of more atoms. The frequency $\omega_j$ will not appear if there are not at least $j+1$ atoms, but once it is present it will not be modified if more atoms are added. This could be advantageous in an experiment where the number of atoms trapped is not known.
\subsection{Other Experimental Challenges}
\label{ssec:other}
A number of experimental considerations could complicate the treatment presented here. Some of these are factors which were consciously discarded in order to make the system more amenable to solutions, like a classical treatment of motion. Others, however, involve additional elements which were never included. 

\begin{figure}[b]
	\includegraphics[width=8.6cm]{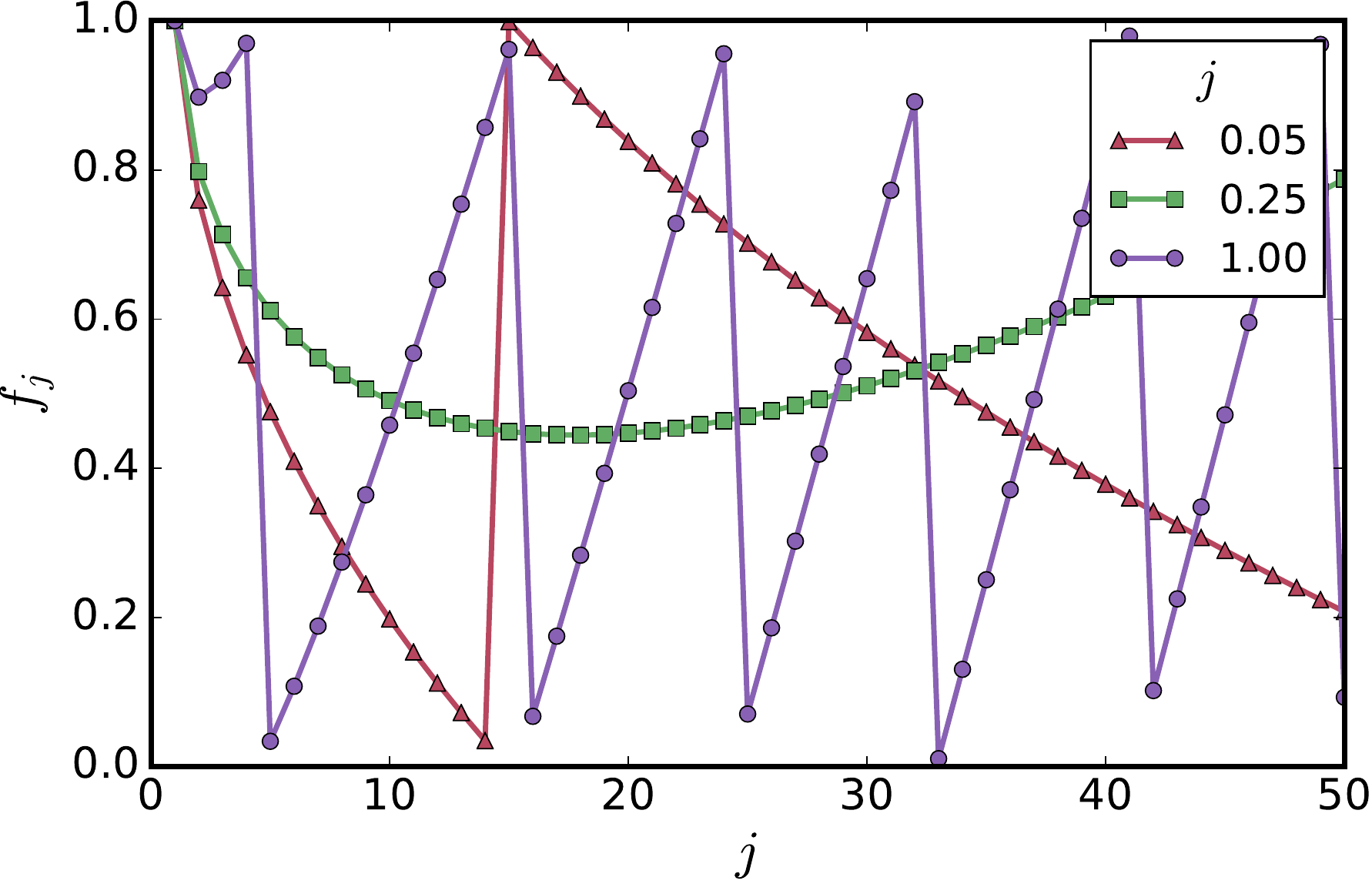}
	\caption{ Spatial configurations of self-organized solutions at several different strengths of $\gamoned$. These calculations were made at $\delta =  1 \gamtot$.}
	\label{fig:gam1Dcompare}
\end{figure}

The precise value of $\gamoned/\gamtot$ could affect the self-organizing behavior of the atomic ensemble. This paper uses $\gamoned = 0.25 \gamtot$, which, while achievable in photonic crystal waveguides, is overly optimistic for nanofiber setups. However, while coupling strengths are an important practical concern for conducting the experiment, they do not have a large effect on the results presented here. As an example, consider the $\chi_r = 1$ case. Far from resonance, the single-particle potentials given in Eq.~\eqref{eqn:singleparticlepotential} apply. Since the spatial period of the sinusoidal potential a particle experiences is given by $k$ and does not depend on decay rates, equation \eqref{eqn:recursive} gives minima at the same places for any $\gamoned$, so the spacings will be the same for all $\gamoned$. Closer to resonance, changing $\gamoned$ does affect the spacing. As shown in \figref{fig:gam1Dcompare}, the effect is a change in lattice spacings. 

Another broad class of concerns would be those that violate the condition of 1D dynamics. The first concern would be that atomic motional modes in the transverse direction could be excited. However, if the main effect of these perturbations is to bring the atoms closer or further from the waveguide, this can be modeled as random noise on the value of $\gamoned$. Figure~\ref{fig:gam1Dcompare} suggests that self-organized configurations will not be sensitive to this effect. These modes could also be resonant with the motional modes shown in \figref{fig:motion}. For nanofiber platforms, this concern also applies to torsional modes \cite{Wuttke2013} which could be resonant with motional modes of atomic ensembles and lead to heating. These mode energies, their couplings to atoms, and the associated timescales will be system-specific, and such particulars are outside the scope of our work.

The experimental procedure required for preparation of a self-organized state has not been discussed. If the desired pump is far from resonance, numerical simulations show that initial positions do not affect the final steady-state configuration. However, in practice, it may be more convenient to release atoms from a longitudinal standing-wave potential trapping them all at $f_j = 1$. Closer to resonance, an experiment could be designed to emulate the procedure of adiabatically adjusting the detuning from the weak-scattering limit to near resonance. In all cases, the methods of this paper could be applied to compute the expected steady state resulting from the particular initial conditions of the experiment. 

Our semi-classical treatment of the atomic motion concerned only expectation values of momenta and positions, and did not address the variance of these operators. However, this is an oversimplification, which led to all atoms having identical momenta. In reality, the incoherent emission will lead to heating of the motional degrees of freedom. A more complicated treatment of the problem could either treat the motion quantum mechanically or implement classical stochastic dynamics.
	Because damping exists in the system, the end result will be a finite temperature \cite{Foot2005}.  In order to evaluate the final achieved temperature, we disregard transverse degrees of freedom and then apply the Einstein diffusion relation to the one-dimensional motion of an atom \cite{Metcalf2003}. In our system this takes the form of:
	\begin{equation}
		m \gamma_p k_B T = D
		\label{eqn:einstein}
	\end{equation}
	Here $D$ is the momentum diffusion coefficient in the system. Focusing on the weak-scattering limit, the rate of momentum diffusion will be given by $\half \left( \hbar k s_0 \right)^2 \Gamma_\mathrm{tot}$ \cite{Metcalf2003},
\begin{equation}
	D = \left( \hbar k \right)^2 \frac{ \abs{\Omega}^2 \gamtot /2 }{ \left( \gamtot /2 \right)^2 + \delta^2 }.
	\label{eqn:diffusion}
\end{equation}
This is the average rate of emission $(s_0^2 \gamtot/2)$ multiplied by the square of the size of the momentum ``kick'' received $(\hbar k)$. Note that in reality the emission of a photon may not provide a full $\hbar k$ of transverse momentum depending on its direction of emission, but to simplify the calculation we assume the worst-case diffusion coefficient. Equations \ref{eqn:einstein} and \ref{eqn:diffusion} can be rearranged to yield:
\begin{equation}
	T = \frac{\hbar \omega_r \gamtot s_0^2}{\gamma_p k_B} \approx \frac{.98 \mathrm{nK} }{\gamma_p / \gamtot}.
\end{equation}
Here we have substituted the physical parameters of the $D_2$ line of neutral cesium and $s_0 = .1$, as elsewhere in the work. To evaluate whether this temperature is problematic for chiral self-organization, we compare the characteristic energy to the minimum trapping energy in Eq.~\eqref{eqn:recursive}. This condition becomes
\begin{align}
	\frac{E_1}{k_B T} =\frac{2 \gamoned}{\omega_r} \left( \gamma_p / \gamtot \right) &\gg 1 \\
	\label{eqn:condition}
\end{align}
For the value of $\gamoned = .25 \gamtot$ we used in our simulations, this would imply $\gamma_p > \omega_r$. This damping rate is impossible for Doppler cooling in the two-level approximation, but is accessible for Sisyphus cooling schemes \cite{Metcalf2003,Dalibard1985}. However, if the two-level approximation is abandoned then there is no need for us to necessarily cool using the same transition which mediates the chiral interaction, meaning that Doppler cooling may still suffice. Note that none of these considerations affect which states are steady states of the equations of motion, but simply give conditions required to achieve these steady states.
\section{Conclusion and Outlook}
\label{sec:conclusion}
We have looked at self-organization in systems with direction-dependent atom-light coupling and found that self-organization survives in a reference frame moving along with the resulting center-of-mass motion. Handling this center-of-mass motion may require damping the atomic motion. In parts of parameter space, chirally self-organized configurations resemble the symmetric case, but in others, especially at full chirality and near resonance, we predict dramatic differences. In the weak-scattering limit the purely chiral case admits an iterative solution which shows a lattice constant slowly increasing with atomic index, while at resonance we observed phase slip configurations at low chiralities which vanish as chirality increases. We have also presented schemes for achieving variable chirality in current optical nanofiber systems and listed expected signatures of the chiral interaction and its characteristic steady-state configurations. In particular, we calculated the optical response and the phonon spectrum of atoms in such systems, which might allow for interrogation via sidebands or parametric heating.

We believe that chiral nanophotonic systems offer many exciting opportunities for quantum optics and atomic physics. Extensions of this paper could include relaxing our assumption of no atomic saturation or low density, adopting more complicated distance dependence of our interactions, and treating the atomic motion in quantum mechanically \cite{Douglas2015}. Since photon-photon interactions mediated by atomic ensembles have attracted great interest, these phenomena could be examined in regimes with asymmetric coupling. This includes electromagnetically induced transparency \cite{LeKien2015} as well as photon bound states \cite{Firstenberg2013,Douglas2015,Maghrebi2015}. The ability of these systems to emulate a spin model means they may be useful for probing the behavior of chiral spin models \cite{Pichler2015,Ramos2014,Stannigel2012,Garttner2015}. Finally, the fact that these one-dimensional ensembles may be useful for photon storage \cite{Sayrin2015} means that a chiral coupling might prove useful for photonic network or integrated photonic circuit applications \cite{Metelmann2015a}.
\section{Acknowledgments}
This work was supported by the NSF PFC at the JQI, AFOSR, NSF QIS, ARO, ARL CDQI, ARO MURI, ERC Starting Grant FOQAL, the MINECO Plan Nacional and Severo Ochoa programs, and Fundacio Privada Cellex. The authors would like to thank Luis Orozco and Steve Rolston for input on experimental considerations, and Christopher Jarzynski, Eugene Polzik, Phillip Schneeweiss, and Ana Asenjo for discussions.

\appendix*
\section{Reduction of Full Hamiltonian to Spin Model}
\label{app:reduce}
In this appendix, we begin from a full atom-field Hamiltonian and obtain the master equation [Eqs.\ \eqref{eqn:master}-\eqref{eqn:chidecay}] governing the atoms.

The full Hamiltonian is given by:
\begin{align}
	H &= \sum_j \frac{p_j^2}{2m} + \hbar \omega \see{j} + i \hbar v \int \dd{z} \left( a_L^\dag (z) \frac{\partial a_L}{\partial z} - a_R^\dag (z) \frac{\partial a_R}{\partial z} \right) \nnb
&- \hbar \sqrt{2 \pi} \int \dd{z}  \bigg[\sum_j \delta \left(z - z_j\right) \left ( \seg{j} \left(\beta_L a_L(z) + \beta_R a_R(z) \right) \right) \nnb &+ h.c. \bigg].
\end{align}
Here the $a_L, a_R$ are the field annihilation operators for left and right propagating fields, respectively, while the $\sigma^j_{\mu, \nu} = \ket{\mu} \bra{\nu}$ are atomic state operators for each atom, where $\mu, \nu \in \left\{ e, g \right\}$ stand for excited or ground states. $\beta_\mathrm{L,R}$ is the atom-field coupling strength for each mode and $v$ is the speed of light in the waveguide. This Hamiltonian does not include the emission into free space, which has jump operator:
\begin{equation}
L_j = \sqrt{\gamma} \sge{j}.
\end{equation}

Formally integrating the equations of motion for $a_R$ and $a_L$, we write the field operators as a sum of the input field and a contribution from the radiating atoms \cite{Chang2012}:
\begin{align}
\label{eq:fields}
a_R(z, t) &= a_{R, \mathrm{in}} ( z- vt) \nnb&+ \frac{i \sqrt{2 \pi} \beta_R}{v} \sum_k \Theta (z - z_k) \sge{j} \left(t - \frac{\abs{z - z_k}}{v} \right), \\
a_L(z, t) &= a_{L, \mathrm{in}} ( z+ vt) \nnb&+ \frac{i \sqrt{2 \pi} \beta_L}{v} \sum_k \Theta (z_k - z) \sge{j} \left(t - \frac{\abs{z - z_k}}{v} \right). \label{eq:fields2}
\end{align}
Here $\Theta$ is the Heaviside theta function. The Heisenberg equation of motion for $\sge{j}$ is
\begin{align}
	\label{eqn:sdot}
\dot{\sigma}^{j}_{ge} &= -i \omega \sge{j} - i\sqrt{2 \pi} \left( \see{j} - \sgg{j} \right) \left[ \beta_L a_L (z_j) + \beta_R a_R (z_j) \right].
\end{align}
Now insert Eqs.~\eqref{eq:fields} and \eqref{eq:fields2} into Eq.~\eqref{eqn:sdot} and make an approximation that $\sge{j}$ is slowly varying. 
Specifically, 
rather than treat the retardation exactly, we take $\sge{j}(t-\epsilon) \approx \sge{j}(t) e^{i \omega_L \epsilon}$, essentially reducing the retardation effect to a relative phase depending on the frequency of light $\omega_L$. This approximation is valid as long as the bandwidth of the dynamics $\Delta \omega$ satisfies $\Delta \omega L/ v \ll 1$, where $L$ is the system size. The equation of motion becomes
\begin{align} \label{eq:sge}
\dot{\sigma}^j_{ge} &= -i \omega \sge{j} -\frac{\gamma}{2} \sge{j} + F(t) + \left( \see{j} - \sgg{j} \right) \times \nnb &\bigg[ -i \left( \sqrt{2 \pi} \beta_L a_{L, \mathrm{in}} (z_j + vt) +  \sqrt{2 \pi} \beta_R a_{R, \mathrm{in}} (z_j - vt) \right) +  \nnb & \frac{2 \pi \beta_L^2}{v} \sum_i \Theta(z_i - z_j) \sge{i} e^{i k \left| z_j - z_i \right|} \nnb &+ \frac{2 \pi \beta_R^2}{v} \sum_i \Theta(z_j - z_i) \sge{i} e^{i k \left| z_j - z_i \right|}   \bigg].
\end{align}
This includes free space decay term and the associated Langevin noise $F(t)$ 
\cite{Gardiner1991}.

If only one atom is present, then there is only one term in each of the two sums in Eq.~\eqref{eq:sge}, and they together contribute $\frac{2\pi (\beta_R^2 +\beta_L^2)}{v} \sge{j}$. The single-atom spontaneous emission rate into the waveguide is 
$\Gamma = \Gamma_L + \Gamma_R = \frac{2 \pi \beta_L^2}{v} + \frac{2 \pi \beta_R^2}{v}$,
which agrees with the symmetric case in the limit $\beta_L = \beta_R$. Removing the single-atom spontaneous emission from the sum and moving to the rotating frame yields the final equation for the evolution of $\sge{j}$:
\begin{align}
\dot{\sigma}^j_{ge} &= \left(i\delta -\frac{\gamma + \Gamma}{2} \right) \sge{j} + \left( \see{j} - \sgg{j} \right) \times \nnb &\bigg[ - i\sqrt{2 \pi} \left( \beta_L a_{L, \mathrm{in}} (z_j + vt) + \beta_R a_{R, \mathrm{in}} (z_j - vt) \right) \nnb &+ \Gamma_L \sum_{i \neq j} \Theta(z_i - z) \sge{i} e^{i k \left| z_j - z_i \right|} \nnb &+ \Gamma_R \sum_{i \neq j} \Theta(z - z_i) \sge{i} e^{i k \left| z_j - z_i \right|}   \bigg].
\end{align}

At this point, in the symmetric case, $\Gamma_L = \Gamma_R$ allows the sums to be combined. In an attempt to duplicate as many features of the symmetric case as possible, we define $\chi = \Gamma_R - \Gamma_L$. Then the equation of motion can be rearranged to contain one complete and one incomplete sum:
\begin{align}
	\label{eqn:hbergeom}
\dot{\sigma}^j_{ge} &= \left(i\delta -\frac{\gamma + \Gamma}{2} \right) \sge{j} +  \left( \see{j} - \sgg{j} \right)\times \nnb & \bigg[ - i\sqrt{2 \pi} \left( \beta_L a_{L, \mathrm{in}} (z_j - vt) + \beta_R a_{R, \mathrm{in}} (z_j + vt) \right) \nnb &+ 
\Gamma_L \sum_{i \neq j} \sge{i} e^{i k \left| z_j - z_i \right| } \nnb &+ \chi \sum_{i\neq j} \Theta(z_j - z_i) \sge{i} e^{i k \left|z_j - z_i \right|} \bigg].
\end{align}
In the absence of input fields,
these are precisely the dynamics yielded by the Hamiltonian, Eq.~\eqref{eqn:hamiltonian}, as well as the jump operators of Eqs. \eqref{eqn:3Ddecay} - \eqref{eqn:chidecay}.

\bibliography{./library}{}

\end{document}